\documentclass[12pt]{iopart}
\usepackage{graphicx}
\usepackage{iopams}

\begin{document}

\title{Laser-assisted spin-polarized transport in graphene tunnel junctions}
\author {Kai-He Ding$^{1,2}$, Zhen-Gang Zhu$^{1}$, and Jamal Berakdar$^{1}$}
\address{$^1$Institut f\"{u}r Physik,
Martin-Luther-Universit\"{a}t Halle-Wittenberg,  06099 Halle
(Saale), Germany} 
\address{$^2$Department of Physics and Electronic Science,
 Changsha University of Science and Technology,
 Changsha,410076, China}

\date{\today }

\begin{abstract}
 The Keldysh nonequilibrium Green's function
method is utilized to study theoretically the spin polarized
transport through a graphene spin valve  irradiated by a
monochromatic laser field. It is found that the bias dependence of
the differential conductance exhibits two peaks
corresponding to the resonant tunneling through the photon-assisted
subbands. A zero value plateau in the differential conductance
appears symmetrically on the zero bias due to the dynamical gap
opened by the radiation field, and its width can be tuned by
changing the radiation intensity  and frequency, which leads to the
shift of the resonant peaks in the differential conductance as well. We
demonstrate that the shift of the resonant peaks in the
differential conductance with the radiation intensity exhibits
different features from that with the radiation frequency. We also
show numerically the dependencies of the radiation and spin valve
effects on the parameters of the external fields and those of the
electrodes. We find that the combined effects of the radiation
field, the graphene, and the spin valve properties bring about a
plateau in the tunnel magnetoresistance, and the width of this
plateau can be changed by scanning the radiation field strength
and/or the frequency.
\end{abstract}

 \maketitle
\section{Introduction}
Electromagnetic radiation was demonstrated to be  a powerful tool for the exploration  and
modification of  materials properties.
For example,  in conventional semiconducting
nanostructures a variety of applications have been established  such as the various forms of  photo electronic
devices (e.g., the radiation-controlled field-effect transistors,
photodiodes, and the light-emitting
diodes  \cite{xdnat2001,msmjap2006,mfnl2003,yoapl2004,jwsci2001}).
On the other hand, graphene with its many
unusual physical properties (to mention but a few, the half integer quantum Hall
effect \cite{ksn2005}, the Klein tunneling \cite{mikn2006} and the
conductance properties \cite{geim})  offers a new platform to
explore  radiation effects. The influence of electromagnetic fields on
  graphene has been recently a subject of intense research.
Particularly interesting examples are the linear response to
 and the
 frequency dependence of the
conductivity \cite{peres,falkepjb2007,gusyprb2006}, the
photon-assisted transport \cite{trauprb2007,dkh}, the microwave and
far-infrared response \cite{gusyprb2007,falkprb2007,aberprb2007},
 the plasmon
spectrum \cite{hwangprb2007,wunsnjp2006,apalijm2007}, as well as
the nonlinear response to the electromagnetic
radiation \cite{vafeprl2006,mikhprl2006,mikjpcm2008,mosk}.  The potential applications
of graphene  in terahertz electronics were pointed out in Refs.\cite{apl2011}.

 The resonant interaction between  graphene and an
electromagnetic field may open a dynamical gap in the
quasiparticle spectrum of graphene \cite{mvfprl2007}, which leads
to a strong suppression of  the quasiparticle transmission through
a graphene p-n junction. However, a directed current without
applying any dc bias voltage can be generated in certain
conditions as a result of inelastic quasiparticle tunneling
assisted by one- or two-photon absorptions \cite{svsprb2008}.
Another way to generate a direct current is to use shaped, timely
tuned electromagnetic pulses \cite{mosk} which also may generate a
valley current.
 Therefore, it seems  possible to control
the transport properties of diverse graphene tunneling structures
by the variation of the strength and the frequency of the external
radiation fields. On the other hand, the potential of graphene
for spin-dependent transport (spintronic) applications is well documented by now
\cite{hill2006ieee,tombros2007nature,cho2007apl,ohishi2007apl,wang2008prb,ding2009prb,maassen,chenjpcm2009,ding}.
 Motivated by these facts, in this work, we study
theoretically the spin polarized transport through a graphene-spin-valve device in the presence
 of  a monochromatic laser field. The method is based on the standard Keldysh nonequilibrium
Green's function approach, as described in Refs.\cite{hh} and
\cite{jr}. It is found that the bias dependence of the
differential conductance exhibits two implicit peaks corresponding
the resonant tunneling through the photon-induced subbands. The
resonant interaction of the quasiparticle in graphene with the
radiation field turns on a dynamical gap in the quasiparticle
spectrum. When the bias voltage lies inside this gap region, the
differential conductance displays a zero value plateau situated
symmetrically on the zero bias. The width of this plateau can be
tuned by changing radiation intensity and frequency, which also
causes a shift of the resonant peaks in the differential
conductance. The tunnel magnetoresistance (TMR) versus the bias
voltage exhibits a plateau around the zero bias voltage, and its
width can be controlled by the radiation field strength and/or the
frequency.

\section{Theoretical model}
We consider a spin valve device consisting of an extended graphene sheet (that defines the $x-y$ plane)
 contacted
by two ferromagnetic electrodes.
 A gate voltage $V_g$ applied to graphene shifts
the Dirac point away from the zero energy. The two electrodes are
voltage-biased with respect to each other with a bias $V$.
 The electrical current flows
in the $x$ direction. Additionally, we assume that a laser field
is  irradiating  homogenously the structure. Within the metallic
electrodes we assume that the field is shielded and ignore thus its
effect on the electrodes. The laser field is monochromatic and is
linearly polarized along the $y$ direction. The low energy
graphene Hamiltonian around the Dirac points has the minimal
coupling form
\begin{equation}
H_{G}=v_{F}\bsigma\bdot\mathbf{k}-ev_{F}A(t)\sigma_y,
\label{hg1}
\end{equation}
where  $\mathbf{k}$ is the operator of the electron momentum in
the graphene plane, $e$ is the electron charge, $\boldsymbol{\sigma}$ is the vector
build out of the Pauli
matrices in the sublattice space, and $A$ is the laser's vector potential. In the tight binding
description \cite{castrocond}, $v_F=\frac{3}{2}at_g$ with $t_g$
being the nearest neighbor hopping energy, and $a$ being the
carbon-carbon distance.
The vector potential $A(t)$ is taken as
\begin{equation}
A(t)=\frac{E_0}{\omega_0}\sin\omega_0t,
\end{equation}
where $\omega_0$ is the frequency of the radiation field, and
$E_0$ is its amplitude. Diagonalizing the
Hamiltonian (\ref{hg1}) in the absence of the radiation field, one
 finds the eigenvalues
\begin{equation}
\epsilon_{s\mathbf{k}}=sv_F|\mathbf{k}|+V_g
\end{equation}
with $s=\pm$ denoting the band index, and the eigenstates
\begin{equation}
\psi_{s\mathbf{k}}(\mathbf{r})=\frac{1}{\sqrt{\Omega}}e^{i\mathbf{k}\cdot\mathbf{r}}u_{\mathbf{k}}^s,
\end{equation}
 where $\Omega$ is the volume of the system, and
\begin{equation}
u_{\mathbf{k}}^s =\frac{\sqrt{2}}{2}\left(
\begin{array}{c}
1\\
se^{i\phi(\mathbf{k})}
\end{array}\right),\quad \tan\phi(\mathbf{k})=\frac{k_y}{k_x} .
\label{eigst}
\end{equation}
We  introduce the field operators
 \begin{equation}
\Psi_\tau(\mathbf{r},t)=\sum\limits_{\mathbf{k},s}a_{s\mathbf{k}\tau}(t)\psi_{s\mathbf{k}}(\mathbf{r}),
\end{equation}
where $a_{s\mathbf{k}\tau}$ is the usual annihilation  operator for an
electron in the band $s$, with the momentum $\mathbf{k}$ and the spin $\tau$,
and then express the Hamiltonian (\ref{hg1}) in the second quantized
form as
\begin{equation}
\begin{array}{cll}
H_G &=&
\sum\limits_{\mathbf{k}s\tau}\epsilon_{s\mathbf{k}}a_{s\mathbf{k}\tau}^\dag
a_{s\mathbf{k}\tau}- ev_FA(t)
\sum\limits_{\mathbf{k}ss'\tau}d_{ss',\mathbf{k}}
a_{s\mathbf{k}\tau}^\dag a_{s'\mathbf{k}\tau},

\end{array}
\end{equation}
where
\begin{equation}
d_{ss',\mathbf{k}}=\frac{i}{2}(se^{-i\phi(\mathbf{k})}-s'e^{i\phi(\mathbf{k})}).
\end{equation}
Applying the rotating wave approximation \cite{ccg}, we neglect the
energy nonconserving terms.  Accounting for the coupling
between the graphene and the two ferromagnetic electrodes we write for the Hamiltonian of
the complete tunnel junction  in the presence of the laser
\begin{equation}
H=H_G+H_L+H_R+H_T,
\end{equation}
where
\begin{equation}
\begin{array}{cll}
H_G

&=&
\sum\limits_{\mathbf{k}s\tau}\epsilon_{s\mathbf{k}}a_{s\mathbf{k}\tau}^\dag
a_{s\mathbf{k}\tau}\\
&&- ev_F \sum\limits_{\mathbf{k}\tau}\left\{A_0^*e^{-i\omega_0 t}
d_{+-,\mathbf{k}} a_{+\mathbf{k}\tau}^\dag
a_{-\mathbf{k}\tau}\right.\\
&&\left. +A_0e^{i\omega_0 t} d_{-+,\mathbf{k}}
a_{-\mathbf{k}\tau}^\dag a_{+\mathbf{k}\tau}\right\}

\end{array}
\end{equation}
with $A_0=-i\frac{E_0}{2\omega_0}$, and
\begin{equation}
H_\lambda=\sum\limits_{\mathbf{q}\tau}
\varepsilon_{\mathbf{q}\lambda \tau} c_{\mathbf{q}\lambda
\tau}^\dag c_{\mathbf{q}\lambda \tau},\ \ \lambda=L,R,\label{hlab}
\end{equation}
\begin{equation}
H_T=\frac{1}{\sqrt{N}}\sum\limits_{\mathbf{qk}\lambda\tau
s}\left(T_{\mathbf{k}\lambda\mathbf{q}}c_{\mathbf{q}\lambda
\tau}^\dag a_{s\mathbf{k}\tau}+ H.c.\right).\label{ht}
\end{equation}
Equations (\ref{hlab}) and (\ref{ht}) describe respectively the
$\lambda$ electrode and the coupling between the graphene and the
electrodes. $\varepsilon_{\mathbf{q}\lambda\tau}$ is the single
electron energy,
$c_{\mathbf{q}\lambda\tau}^\dag(c_{\mathbf{q}\lambda\tau})$ is the
usual creation (annihilation) operator for an electron with the
momentum $\mathbf{q}$ and the spin $\tau$ in the $\lambda$
electrode; $N$ is the number of sites on the sublattice.

By introducing a unitary transformation
\begin{equation}
U=\exp\left[-i\frac{\omega_0
t}{2}\sum\limits_{\mathbf{k}\tau}(a_{+\mathbf{k}\tau}^\dag
a_{+\mathbf{k}\tau}-a_{-\mathbf{k}\tau}^\dag
a_{-\mathbf{k}\tau})\right],
\end{equation}
we redefine the Hamiltonian of the system in the rotating
reference as
\begin{equation}
\begin{array}{cll}
\widetilde{H}&=&U^{-1}HU+i\frac{d U^{-1}}{dt}U\\

&=&\sum\limits_{\mathbf{k}\tau}\widetilde{\epsilon}_{s\mathbf{k}}a_{s\mathbf{k}\tau}^\dag
a_{s\mathbf{k}\tau}+ \sum\limits_{\mathbf{k}\tau}\Delta\left(
a_{+\mathbf{k}\tau}^\dag a_{-\mathbf{k}\tau}

+ a_{-\mathbf{k}\tau}^\dag
a_{+\mathbf{k}\tau}\right)\\

&+&\sum\limits_{\mathbf{q}\lambda \tau}
\varepsilon_{\mathbf{q}\lambda \tau} c_{\mathbf{q}\lambda
\tau}^\dag c_{\mathbf{q}\lambda
\tau}+\frac{1}{\sqrt{N}}\sum\limits_{\mathbf{k q}\lambda
s\tau}\left(T_{\mathbf{k}\lambda s\mathbf{q}}
 (t)c_{\mathbf{q}\lambda \tau}^\dag
a_{s\mathbf{k}\tau}+ H.c.\right),\\
\end{array}\label{finham}
\end{equation}
where $$\Delta=\frac{ev_FE_0}{2\omega_0},\,
T_{\mathbf{k}\lambda s\mathbf{q}}(t)=T_{\mathbf{k}\lambda
\mathbf{q}}e^\frac{-is\omega_0 t}{2}$$ and
$$\widetilde{\epsilon}_{s\mathbf{k}}=\epsilon_{s\mathbf{k}}-s\omega_0/2.$$
In the calculation of equation (\ref{finham}), we have assumed  (as in Refs. \cite{mvfprl2007,svsprb2008,xzprb2010})
that the most important contributions stem  from an almost one
dimensional electron motion $(k_x\gg k_y)$ in the interaction with
the radiation field.

The electric current of the system can be calculated from the time
evolution of the occupation number operator of the left electrode,
\begin{equation}
I_L=e\langle\dot{ \mathcal{N}_L}\rangle=\frac{ie}{\hbar}\langle
[\widetilde{H},\mathcal{N}_L]\rangle ,\mbox{ where }
\mathcal{N}_L=\sum\limits_{\mathbf{q}\tau}c_{\mathbf{q}L\tau}^\dag
c_{\mathbf{q}L\tau}. \label{jeql}
\end{equation}
 Using the nonequilibrium Green's function
method, equation  (\ref{jeql}) can be further expressed as
\begin{equation}
\begin{array}{cll}
I_L &=&-\frac{ie}{\hbar}\sum\limits_{ ss'} \int
dt_1\int\frac{d\varepsilon}{2\pi}\rm{Tr}
\{[(\mathcal{G}_{s,s'}^{r}(t,t_1)-\mathcal{G}_{s,s'}^{a}(t,t_1))
f_{L}(\varepsilon)\\
&&+\mathcal{G}_{s,s'}^{<}(t,t_1)]\Gamma_{L}\}
e^{-i\varepsilon(t_1-t)}e^{-\frac{is\omega_0
t}{2}}e^{\frac{is'\omega_0 t_1}{2}},
\end{array}\label{curi}
\end{equation}
where Tr is the trace in the spin space,
$f_{\lambda}(\varepsilon)$ is the Fermi distribution,
$\mathcal{G}_{s,s'}^{r(a)}(t,t_1)=\frac{1}{N}\sum\limits_{\mathbf{kk}'}G_{\mathbf{k}s,\mathbf{k}'s'}^{r(a)}(t,t_1)$,
and
$\mathcal{G}_{s,s'}^{<}(t,t_1)=\frac{1}{N}\sum\limits_{\mathbf{kk}'}G_{\mathbf{k}s,\mathbf{k}'s'}^{<}(t,t_1)$
are $2\times 2$ matrices denoting the retarded (advanced) Green's
function and the lesser Green's function, respectively. In the
calculation of equation  (\ref{curi}), we assume that the dominant
contributions to the tunneling stem from the electrons near the Fermi
level, and hence the linewidth function to be independent of
$\mathbf{k}$. Thus, we have
\begin{equation}
\Gamma_{\lambda }=\left(
\begin{array}{cc}
\Gamma_{\lambda }^\uparrow & 0 \\
0 & \Gamma_{\lambda }^\downarrow%
\end{array}%
\right)
\end{equation}
with $\Gamma_\lambda^\tau =2\pi\sum\limits_\mathbf{q}
T_{\mathbf{k}\lambda\mathbf{q}}^*T_{\mathbf{k'}\lambda\mathbf{q}}
\delta(\varepsilon-\varepsilon_{\mathbf{q}\lambda\tau})$. In order
to solve equation  (\ref{curi}), we need to calculate the Green's
functions $\mathcal{G}_{ss'}^{\tau\tau',r}(t,t') $ and
$\mathcal{G}_{ss'}^{\tau\tau',<}(t,t')$. Using the equation of
motion method, we get
\begin{equation}
\begin{array}{lll}
\mathcal{G}^{\tau\tau',r}(t,t')

=\delta_{\tau\tau'}g^{\tau\tau',r}(t-t')\\
+
 \int dt_1dt_2 g^{\tau\tau,r}(t-t_1)\Sigma^{\tau,r}(t_1,t_2)

\mathcal{G}^{\tau\tau',r}(t_2,t'),
\end{array}\label{grematnm}
\end{equation}
where
\begin{equation}
\Sigma_{ss'}^{\tau,r}(t,t')=\int\frac{d\varepsilon}{2\pi}\Sigma_0^{\tau,r}e^{-i\varepsilon(t-t')}
e^{i\frac{s\omega_0 t}{2}}e^{-i\frac{s'\omega_0 t'}{2}}
%
\end{equation}
with
$\Sigma_0^{\tau,r}=-\frac{i}{2}(\Gamma_L^\tau+\Gamma_R^\tau)$, and
$g^{\tau\tau',r}(t)$ is the retarded Green's function of graphene
without the coupling of the electrodes, and can be obtained by a
straightforward calculation. The detailed expressions are given in the
Appendix. $\mathcal{G}^{\tau\tau',<}(t,t')$ is related to
$\mathcal{G}^{\tau\tau',r}(t,t')$ through the Keldysh equation
\begin{equation}
\begin{array}{cll}
\mathcal{G}^{\tau\tau,<}(t,t')&=&\int dt_1dt_2
G^{\tau\tau,r}(t,t_1)\Sigma^{\tau,<}(t_1,t_2)G^{\tau\tau,a}(t_2,t'),
\end{array}\label{kel}
\end{equation}
where
\begin{equation}
\begin{array}{cll}
\Sigma_{ss'}^{\tau,<}(t,t')&=&i\int\frac{d\varepsilon}{2\pi}[\Gamma_L^\tau
f_L(\varepsilon)+\Gamma_R^\tau
f_R(\varepsilon)]\\

&&\times e^{-i\varepsilon(t-t')}e^{i\frac{s\omega_0
t}{2}}e^{-i\frac{s'\omega_0 t'}{2}}.
\end{array}
\end{equation}
Notice that the Green's functions in equations  (\ref{grematnm}) and
(\ref{kel}) do not depend only on the difference of the two time
variables, thus one should take a generalized Fourier expansion
as \cite{levyprb}
\begin{equation}
\mathcal{G}(t,t')=\frac{1}{2\pi}\sum\limits_n\int d\varepsilon
e^{-i\varepsilon
t}e^{i(\varepsilon+n\omega_0/2)t'}\mathcal{G}(\varepsilon,\varepsilon+n\omega_0/2).\label{getran}
\end{equation}
Hereafter we shall use the simple notation
$$\mathcal{G}_{nm}(\varepsilon)=\mathcal{G}(\varepsilon+n\omega_0/2,\varepsilon+m\omega_0/2).$$
Evidently, the different Fourier components satisfy the relation
$$\mathcal{G}_{nm}(\varepsilon)=\mathcal{G}_{n-m,0}(\varepsilon+m\omega_0/2).$$
From equation  (\ref{grematnm}), the Fourier components of the Green's
function can be expressed as
\begin{equation}
\begin{array}{cll}
 \mathcal{G}_{m,n}^{\tau\tau,r}&=&g_{mn}^{\tau\tau,r}\\
&&
+\widetilde{\epsilon}_m\mathcal{G}_{m,n}^{\tau\tau,r}+V_{m,m-2}\mathcal{G}_{m-2,n}^{\tau\tau,r}
 +V_{m,m+2}\mathcal{G}_{m+2,n}^{\tau\tau,r},
\end{array}\label{gremx1m}
\end{equation}
where
$\widetilde{\epsilon}_m=g_{mn}^{\tau\tau,r}\Sigma_0^{\tau,r}$, and

$$
\begin{array}{cll}
V_{m,m-2}&=& \left(
\begin{array}{cc}
 g_{mm;+-}^{\tau\tau,r}(\varepsilon) \Sigma_0^{\tau,r}& 0\\
  g_{mm;--}^{\tau\tau,r}(\varepsilon) \Sigma_0^{\tau,r}&0\\
 \end{array}\right),
\\
V_{m,m+2}&=& \left(
\begin{array}{cc}
 0&g_{mm;++}^{\tau\tau,r}(\varepsilon) \Sigma_0^{\tau,r}\\
 0& g_{mm;-+}^{\tau\tau,r}(\varepsilon) \Sigma_0^{\tau,r}\\
 \end{array}\right),
\end{array}\label{gremtx1}
$$
with $g_{mn,ss'}^{\tau\tau,r}(\varepsilon)=\delta_{mn}g_{ss'}^{\tau\tau,r}(\varepsilon+m\omega_0/2)$. Note that equation  (\ref{gremx1m}) is formally equivalent to the ones
describing the motion of electrons in a tight-binding linear chain
with site energies and the nearest-neighbor coupling. Its solution
can be derived by using the conventional recursive
technique \cite{levyprb}.

Let $\Gamma_{L}^{\tau}=\gamma\Gamma_{R}^{\tau}$, and $N(t)$ be the
occupation of the central graphene. By using the continuity
equation,
$$\frac{dN(t)}{dt}=I_R(t)+I_L(t),$$
we finally obtain the time-averaged current
\begin{equation}
\begin{array}{cll}
\langle I\rangle

&=&-\frac{\gamma}{1+\gamma}\frac{ie}{\hbar}\sum\limits_{
ss'\tau}\int\frac{d\varepsilon}{2\pi}
  [\mathcal{G}_{-s,-s';ss'}^{\tau\tau,r}(\varepsilon)
-\mathcal{G}_{-s,-s';ss'}^{\tau\tau,a}(\varepsilon)]
\Gamma_{R}^{\tau}[f_{L}(\varepsilon)-f_{R}(\varepsilon)],
\end{array}\label{fjf}
\end{equation}
where $\mathcal{G}_{mn;ss'}^{\tau\tau,r}(\varepsilon)$ is the
matrix elements of $\mathcal{G}_{mn}^{\tau\tau,r}(\varepsilon)$.
In equation  (\ref{fjf}), we further assume a symmetrical voltage
division: $\mu_{L;R} = E_F\pm 1/2eV$, and put $E_F = 0$ in the
numerical calculations. The expression (\ref{fjf}) is the central
result of this paper, and allows one to describe the
spin-polarized transport within the rotating wave approximation.
WE note the  difference to our previous study \cite{dkh} in
which we studied the influence of a time-dependent chemical
potential (realized as an oscillating bias voltage). In the
present study
 the radiation field, coupled to the carrier as  a vector
potential, induces
a dynamical gap in the quasiparticle spectrum. This, in turn,  leads
to  changes in the transport properties and TMR, as shown
below.
 The TMR is  deduced according to the
  conventional definition
\begin{equation}
\rm{TMR}=\frac{I(0)-I(\pi)}{I(0)},\label{tmr}
\end{equation}
where $I(0(\pi))$ is the time-averaged current flowing through the
system in the parallel (antiparallel) configuration.

\section{Numerical analysis}
Based on the above analytical expressions, we present the
numerical calculations for the spin polarized transport through
the system. Firstly, we assume the linewidth function
$\Gamma_\lambda^\tau(\varepsilon)$ to be independent of the energy
within the wide band approximation, and the two electrodes are
made of the same material.  Introducing the degree of the spin
polarizations of the left and the right electrodes $p_L=p_R=p$, we
can write
$\Gamma_L^{\uparrow\downarrow}=\Gamma_R^{\uparrow\downarrow}=\Gamma_0(1\pm
p)$ where $\Gamma_0$ describes the coupling between the graphene
and the electrodes in the absence of an internal magnetization. We
set the temperature to zero and  $t_g\approx 2.8$ eV in our
calculation. In addition, we choose the frequency
$\hbar\omega_0\leq 0.06t_g$, and the voltage $eV< 1.4t_g$ so that
the Dirac-type behaviour is maintained  in this energy regime
\cite{castrocond}.

\begin{figure}
\center
\includegraphics[width=0.6\columnwidth ]{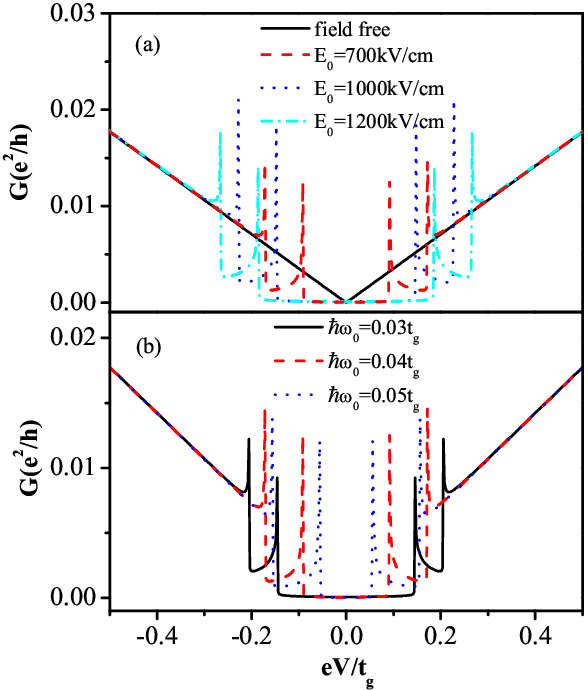}
\caption{(color online)The bias dependence of the differential
conductance $G$ for different radiation strength $E_0$  at
 $\hbar\omega_0=0.04t_g$ (a),  and for different
radiation frequency $\omega_0$ at $E_0=700kV/cm$ (b). The other
parameters are taken as $P=0.4$, $\Gamma_0=0.05t_g$, $V_g=0$ and
$D=3t_g$. The inset in (a) shows the differential conductance in
the absence of the radiation field.}\label{fig1}
\end{figure}

 Fig.\ref{fig1}(a) shows the
the differential conductance $G=d\langle I\rangle/dV$ as a
function of the bias voltage for different radiation strengths in
the parallel configuration of the two electrodes. In the absence
of the radiation field, the differential conductance
 exhibits  a linear dependence on the bias voltage (see Fig.\ref{fig1}(a)).
 The graphene spin valve device is sensitive to the electromagnetic radiation.
 When the radiation field is applied, some characteristic features are presented explicitly in the transport.
 Firstly, close to the zero bias voltage, a salient zero value
region appears symmetrically in the differential conductance, and
the width of the region becomes larger with increasing the
radiation strength. This is because the resonant interaction
between the graphene and the external radiation field opens a
dynamical gap in the spectrum of the quasiparticle in the
graphene. When the bias voltage lies inside this gap, the graphene
can be viewed as an insulator, thus the electron transport through
the graphene is suppressed. Secondly, the
differential conductance as a function of the bias voltage
exhibits two peaks corresponding to the resonant tunneling via the photon-induced sidebands.
This behavior is different to the
case of the presence of an oscillating  gate voltage in graphene sheet (a successive peaks are observed )\cite{dkh}.
The reason for this difference is that  the sidebands in the latter case
 are produced due to the modulation of each  quasiparticle
level by the gate voltage; while in former case they are
caused by the electron scattering between the sidebands. When the
electrons are injected  from the electrode to the graphene, the
radiation field results in   multiple  electron scattering events and
 two resonant
points  within our approximation \cite{svsprb2008}; for higher intensities that may result
in further structures one has to go beyond the rotating wave approximation.
When the bias windows cross the resonant
points, the resonant tunnelings happen leading to the appearance
of two peaks. In addition, for the case of the gate voltage, when
changing the amplitude of the gate voltage, the positions of the
peaks are fixed. In the present case, with increasing the radiation
strength, the positions of the peaks shift in the direction away
from the origin. This is due to the increase of the dynamical gap
lifting the photon-assisted subbands.
 Fig.\ref{fig1}(b) shows the bias dependence of
the differential conductance for the different frequency
$\omega_0$ in the parallel configuration of the electrodes. With
increasing the frequency, the dynamical gap diminishes leading to
the decrease of the width of zero value plateau and  the shift of
the peaks towards the origin. Especially, one can find that the
amplitude of the peaks change non-monotonously
 with the radiation intensity and the frequency.
The reason is that the decrease of the frequency or the increase
of the radiation intensity enhances the gap, which lifts the
photon-induced subbands, and causes the increase of the density of
state for the subbands,thus enhancing the electrons tunneling
through the graphene. 

\begin{figure}
\center
\includegraphics[width=0.6\columnwidth ]{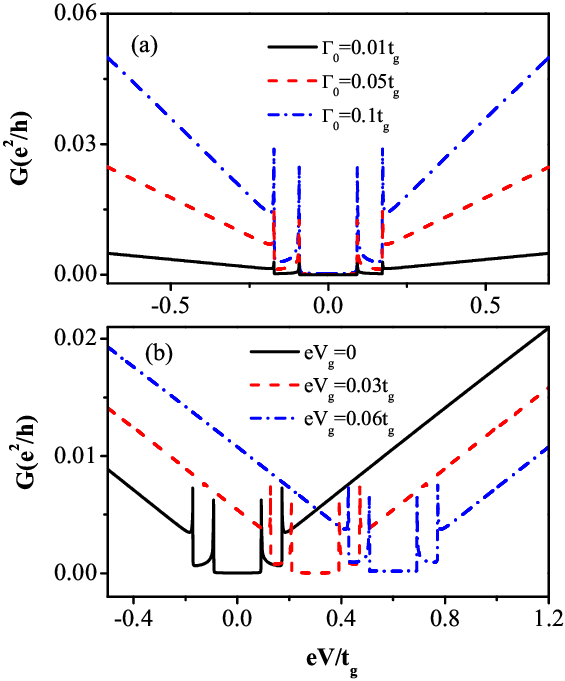}
\caption{(color online)The differential conductance $G$ versus
 the bias voltage (a) for coupling strength $\Gamma_0$ at $V_g=0$, and (b) for different gate voltages $V_g$ at $\Gamma_0=0.05t_g$.
  The other parameters are taken the same as those of
Fig.\ref{fig1}.}\label{fig2}
\end{figure}

Fig.\ref{fig2}(a) shows the differential conductance as a
function of the bias voltage for the different coupling strength
$\Gamma_0$ between the graphene and the electrodes in the parallel
configuration of the electrodes. With increasing the coupling
strength, the electrons tunnel more easily from the electrode to
the graphene, thus leading to the rise of the differential
conductance with $\Gamma_0$. While the zero value plateau in the
differential conductance remains the same because
 the coupling between the electrode and the graphene has no
 influence on the dynamical gap in the quasiparticle spectrum of the graphene.
 Figure \ref{fig2}(b) illustrates  the bias dependence of the differential
conductance for different gate voltages. It shows that the
oscillation peaks and the zero value region shift towards the
positive direction of the bias voltage.

\begin{figure}
\center
\includegraphics[width=0.6\columnwidth ]{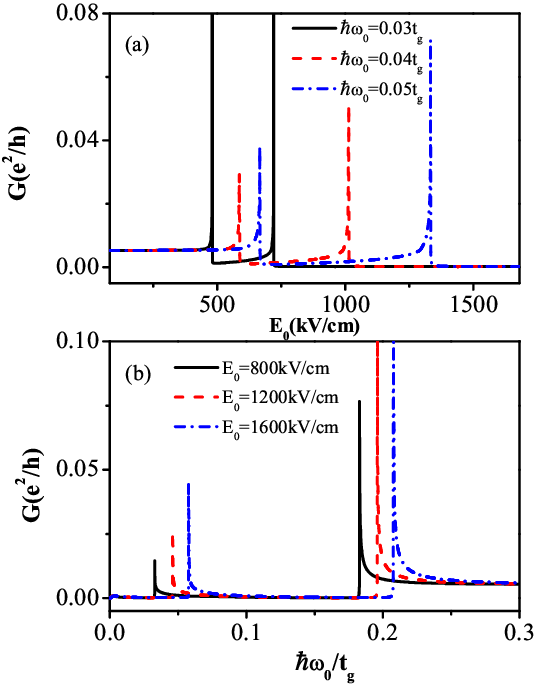}
\caption{(color online)The differential conductance $G$ as a
function of the radiation field strength for  different
frequencies in (a), and as a function of the frequency for different
radiation strengths in (b). In both cases we consider   the parallel
configuration of the electrodes' magnetizations at $eV=0.15t_g$.
The other parameters are taken the same as those of
Fig.\ref{fig1}.}\label{fig3}
\end{figure}

\begin{figure}
\center
\includegraphics[width=0.6\columnwidth ]{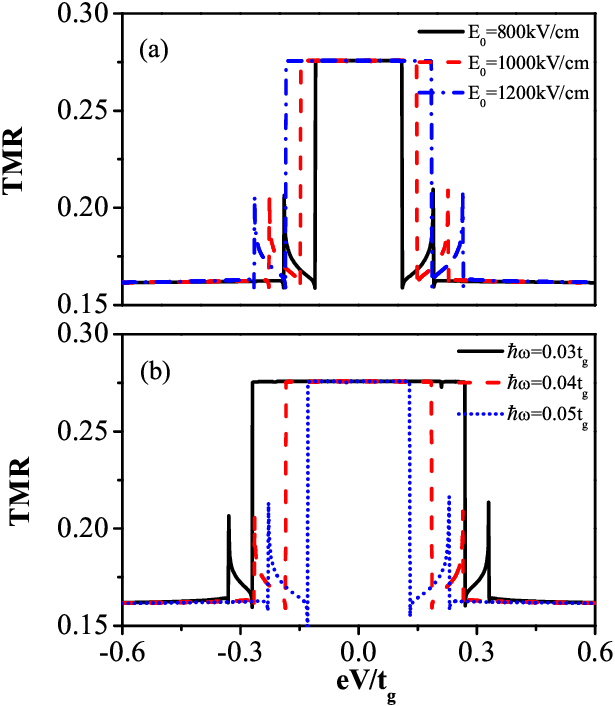}
\caption{(color online) The  TMR versus the bias voltage for
different radiation field strength $E_0$ at
$\hbar\omega_0=0.04t_g$(a), and for different radiation
frequencies $\omega_0$ at $E_0=1200$ kV/cm in (b). The other parameters
are taken the same as those of Fig.\ref{fig1}.}\label{fig4}
\end{figure}

\begin{figure}[tbh]
\center
\includegraphics[width=0.6\columnwidth ]{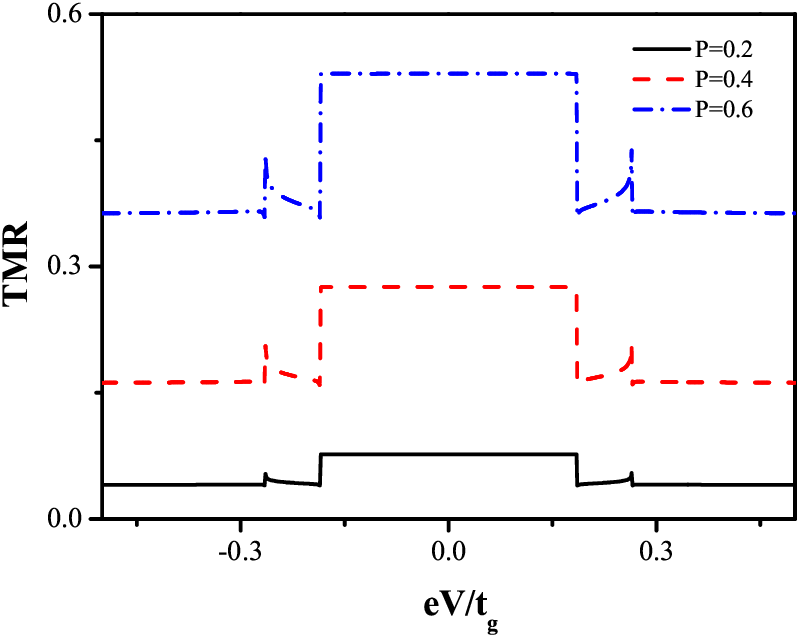}
\caption{(color online) The  TMR versus the bias voltage for a
different polarization $p$ at $\hbar\omega_0=0.04t_g$ and
$E_0=1200$ kV/cm. The other parameters are taken the same as those
of Fig.\ref{fig1}.}\label{fig5}
\end{figure}

The differential conductance as a function of the radiation
intensity for the different radiation frequency in the parallel
configuration of the electrodes is shown in Fig. \ref{fig3}(a).
Two resonant peaks via the photon-induced sidebands can be distinctly observed.
The dynamical gap in the quasiparticle spectrum of the graphene becomes larger with increasing the radiation strength, up-shifting the subbands induced by the radiation field. When the subbands fall into the bias windows,  resonant tunneling occurs, manifested as the conductance peaks.
For radiation with higher frequencies, the resonant peaks appear at larger $E_{0}$ in accordance with a larger dynamical gap.
When $E_{0}$ is large enough, the differential
conductance tends to a constant value. 
Additionally, it is found that the interval between the
resonant peaks 
far away from the origin is larger than that close to the origin.
This behavior can be understood from  Fig.\ref{fig1}(b). The
resonant points at  high energies are close to each other. Thus, a
small difference of the dynamical gap can make them  cross
successively the bias windows, while for the resonant points at
the low energy, a larger one is required to shift them for the
occurrence of the resonant tunneling.
Fig.\ref{fig3}(b) shows the
dependence of the differential conductance on the radiation
frequency $\omega_0$ for  different radiation intensities in the
parallel electrodes configuration. It is clearly seen that when
$\omega_0$ approaches to zero, the differential conductance tends
to vanish owing to the large dynamical gap. With increasing the
frequency, the dynamical gap shrinks, which causes the shift of
the photon-induced subbands, therefore, one can observe that two
resonant peaks emerge successively because of the bias windows
crossing of the subbands which is induced by the external radiation field.
Additionally, the increase of the radiation intensity enlarges the
dynamical gap, thus shifting the resonant peaks in the
differential conductance in the direction away from the origin.
The amount of  the  peaks shift
 at  high
frequencies is the same as at low frequencies. This
is because when changing the radiation strength, the interval of
the resonant points at  high energies is equivalent to that at
the low energies (see Fig.\ref{fig1}(a)). Thus the same variation of
the radiation intensity is needed to achieve the crossing of the
bias windows due to the linear dependence of the
dynamical gap on the radiation field.

The bias dependence of the TMR for the different radiation
strengths and frequencies is shown in Fig. \ref{fig4}. The TMR
as a function of the bias voltage exhibits a sharp peak which is
caused by photon-assisted effects, and a plateau around the zero
bias voltage. We assign this behavior to the combined effects of
the radiation field, the graphene, and the spin valve properties.
In order to further clarify this point, we plot the bias
dependence of the TMR for  different polarizations in Fig. \ref{fig5}.
One can observe that the TMR changes in a nonlinear manner, i.e. the
TMR values near zero bias voltage become larger than those in the
high bias regions. This is because graphene remains insulator-like
in the gap regions, the ballistic spin tunneling enhances the TMR.
Remarkablely, the width of the plateau near the zero bias voltage
can be tuned by the radiation intensity and frequency(see Fig. \ref{fig4}). With increasing the
radiation intensity (or frequency), the width
of the plateau increases (or decreases), which is also related to
the dynamical gap induced by the radiation field.

\section{Summary}
In conclusion, we studied theoretically the spin polarized transport
through a graphene spin valve device assisted by  a linearly polarized, monochromatic
 laser field and in the presence of  a dc bias and a gate voltage. The method is based on the
standard Keldysh nonequilibrium Green's function approach. We find
that the bias dependence of the differential conductance exhibits
two resonant peaks due to the resonant tunneling through the
photon-assisted subbands. The resonant interaction of the
quasiparticle in graphene with the radiation field turns on a
dynamical gap in the quasiparticle spectrum. When the bias voltage
lies inside this gap region, the differential conductance displays
a zero value plateau situated symmetrically to the zero bias. The
value of the dynamical gap depends on the strength and frequency
of the external radiation field, thus the width of this zero value
plateau in the differential conductance can be tuned by changing the
radiation intensity and frequency, which also causes a shift of
the resonant peaks in the differential conductance. We explored
the behavior of the peaks in the differential conductance with
varying radiation field strength and frequency and revealed the
dependence on the spin polarization of the ferromagnetic
electrodes. We also demonstrated that the combined effects of the
radiation field, the graphene, and the spin valve properties bring
about a symmetric plateau at zero bias voltage in the TMR. The
width of this plateau can be varied by changing the parameters of
the radiation field.

\ack The work of K.H.D. is supported by  DAAD (Germany) and by the
National Natural Science Foundation of China (Grant Nos.
10904007), and the construct program of the key discipline in
Hunan Province, China.


\newpage

\appendix

\section{}

In this appendix, we present the analytical results of the Green's
function $g_{ss'}^{\tau\tau',r}(t)$ for the graphene without the
coupling to the electrodes in the absence of the gate voltage.
From the Hamiltonian $H_G$, we find by using the equation of
motion method that
\begin{equation}
\begin{array}{cll}
g_{ss'}^{\tau\tau,r}(t) &=&\int
\frac{d\varepsilon}{2\pi}g_{ss'}^{\tau\tau,r}(\varepsilon)e^{-i\varepsilon
t},
\end{array}
\end{equation}
where $ g
^{\tau\tau,r}(\varepsilon)=\frac{1}{N}\sum\limits_\mathbf{k}g_\mathbf{k}^{\tau\tau,r}(\varepsilon)$
with
\begin{equation}
\begin{array}{cll}
g_\mathbf{k}^{\tau\tau,r}(\varepsilon)

&=&\left(\begin{array}{cc}

\frac{\varepsilon-\widetilde{\epsilon}_{-\mathbf{k}}}
{(\varepsilon-\widetilde{\epsilon}_{+\mathbf{k}})
(\varepsilon-\widetilde{\epsilon}_{-\mathbf{k}})-\Delta^2}

&\frac{\Delta} {(\varepsilon-\widetilde{\epsilon}_{+\mathbf{k}})
(\varepsilon-\widetilde{\epsilon}_{-\mathbf{k}})-\Delta^2}
\\
\frac{\Delta} {(\varepsilon-\widetilde{\epsilon}_{+\mathbf{k}})
(\varepsilon-\widetilde{\epsilon}_{-\mathbf{k}})-\Delta^2}

&\frac{\varepsilon-\widetilde{\epsilon}_{+\mathbf{k}}}
{(\varepsilon-\widetilde{\epsilon}_{+\mathbf{k}})
(\varepsilon-\widetilde{\epsilon}_{-\mathbf{k}})-\Delta^2}

\end{array}\right)
\end{array}
\end{equation}
To solve $g_{ss'}^{\tau\tau,r}(\varepsilon)$, we need to convert
the summation over $\mathbf{k}$ into a integral in two dimensional
momentum space. After a straightforward calculation, we have for
$|\varepsilon|<\Delta$,
\begin{equation}
\begin{array}{lll}
g_{++}^{\tau\tau,r}&=&-\frac{1}{2\pi\rho
v_F^2}\{(\frac{\varepsilon}{2}+\frac{\omega_0}{4})\ln\left|\frac{(D-\omega_0/2)^2+\Delta^2-\varepsilon^2
}{\omega_0^2/4+\Delta^2-\varepsilon^2 }\right|\\

&&-\frac{1}{\sqrt{\Delta^2-\varepsilon^2}}(\Delta^2-\varepsilon^2-\frac{\omega_0}{2}\varepsilon)
[\arctan\frac{D-\omega_0/2}{\sqrt{\Delta^2-\varepsilon^2}}
-\arctan\frac{-\omega_0/2}{\sqrt{\Delta^2-\varepsilon^2}}]+D\}\\
\end{array}\label{app1}
\end{equation}
\begin{equation}
\begin{array}{lll}
g_{--}^{\tau\tau,r} &=&-\frac{1}{2\pi\rho
v_F^2}\{(\frac{\varepsilon}{2}-\frac{\omega_0}{4})\ln\left|\frac{(D-\omega_0/2)^2+\Delta^2-\varepsilon^2
}{\omega_0^2/4+\Delta^2-\varepsilon^2 }\right|\\

&&+\frac{\Delta^2-\varepsilon^2+\varepsilon\omega_0/2}{\sqrt{\Delta^2-\varepsilon^2}}
[\arctan\frac{D-\omega_0/2}{\sqrt{\Delta^2-\varepsilon^2}}
-\arctan\frac{-\omega_0/2}{\sqrt{\Delta^2-\varepsilon^2}}]-D\}\\
\end{array}
\end{equation}
\begin{equation}
\begin{array}{cll}
g_{+-}^{\tau\tau,r}=g_{-+}^{\tau\tau,r} &=&-\frac{1}{2\pi\rho
v_F^2}\left\{\frac{\Delta}{2}\ln\left|\frac{(D-\omega_0/2)^2+\Delta^2-\varepsilon^2
}{\omega_0^2/4+\Delta^2-\varepsilon^2
}\right|\right.\\

&&+\left.\frac{\Delta\omega_0/2}{\sqrt{\Delta^2-\varepsilon^2}}[\arctan\frac{D-\omega_0/2}{\sqrt{\Delta^2-\varepsilon^2}}
-\arctan\frac{-\omega_0/2}{\sqrt{\Delta^2-\varepsilon^2}}]\right\}
\end{array}
\end{equation}
For $|\varepsilon|>\Delta$,

\begin{equation}
\begin{array}{cll}
g_{++}^{\tau\tau,r}

&=&-\frac{1}{4\pi\rho v_F^2}\left[
(\varepsilon+\frac{\omega_0}{2})\ln\left|\frac{(D-\omega_0/2)^2-\varepsilon^2
+\Delta^2}{\varepsilon^2 -\Delta^2-\omega_0^2/4}\right|\right.\\

&&\left.-(\varepsilon\frac{\frac{\omega_0}{2}}{\sqrt{\varepsilon^2
-\Delta^2}}+\sqrt{\varepsilon^2 -\Delta^2})
\ln\left|\frac{(D-\omega_0/2+\sqrt{\varepsilon^2
-\Delta^2})(\omega_0/2+\sqrt{\varepsilon^2
-\Delta^2})}{(D-\omega_0/2-\sqrt{\varepsilon^2
-\Delta^2})(\omega_0/2-\sqrt{\varepsilon^2 -\Delta^2})}\right|+2D
\right]\\

&&+\frac{1}{4\pi\rho v_F^2}i\pi\rm{sgn}(\varepsilon) \{
\theta(\Delta<|\varepsilon|<\sqrt{\frac{\omega_0^2}{4}+\Delta^2})
(\varepsilon-\sqrt{\varepsilon^2
-\Delta^2})(1-\frac{\frac{\omega_0}{2}}{\sqrt{\varepsilon^2 -\Delta^2}})\\

&& -
\theta(\Delta<|\varepsilon|<\sqrt{(D-\frac{\omega_0}{2})^2+\Delta^2})(\varepsilon+\sqrt{\varepsilon^2
-\Delta^2})(1+\frac{\frac{\omega_0}{2}}{\sqrt{\varepsilon^2 -\Delta^2}})\}\\

\end{array}
\end{equation}

\begin{equation}
\begin{array}{cll}
g_{--}^{\tau\tau,r} &=&-\frac{1}{4\pi\rho v_F^2}\left[

(\varepsilon-\frac{\omega_0}{2})\ln\left|\frac{(D-\omega_0/2)^2-\varepsilon^2
+\Delta^2}{\varepsilon^2 -\Delta^2-\omega_0^2/4}\right|\right.\\

&&\left.+(\sqrt{\varepsilon^2
-\Delta^2}-\varepsilon\frac{\frac{\omega_0}{2}}{\sqrt{\varepsilon^2
-\Delta^2}}) \ln\left|\frac{(D-\omega_0/2+\sqrt{\varepsilon^2
-\Delta^2})(\omega_0/2+\sqrt{\varepsilon^2
-\Delta^2})}{(D-\omega_0/2-\sqrt{\varepsilon^2
-\Delta^2})(\omega_0/2-\sqrt{\varepsilon^2 -\Delta^2})}\right|-2D
\right]\\

&&+\frac{1}{4\pi\rho v_F^2}i\pi\rm{sgn}(\varepsilon) \{
\theta(\Delta<|\varepsilon|<\sqrt{\frac{\omega_0^2}{4}+\Delta^2})
(\varepsilon+\sqrt{\varepsilon^2
-\Delta^2})(1-\frac{\frac{\omega_0}{2}}{\sqrt{\varepsilon^2 -\Delta^2}})\\
&& -
\theta(\Delta<|\varepsilon|<\sqrt{(D-\frac{\omega_0}{2})^2+\Delta^2})(\varepsilon-\sqrt{\varepsilon^2
-\Delta^2})(1+\frac{\frac{\omega_0}{2}}{\sqrt{\varepsilon^2 -\Delta^2}})\}\\

\end{array}
\end{equation}
\begin{equation}
\begin{array}{cll}
g_{+-}^{\tau\tau,r}&=&g_{-+}^{\tau\tau,r}\\

&=&-\frac{\Delta}{4\pi\rho v_F^2}
\left[\ln\left|\frac{(D-\omega_0/2)^2-\varepsilon^2
+\Delta^2}{\varepsilon^2 -\Delta^2-\omega_0^2/4}\right|

-\frac{\frac{\omega_0}{2} }{\sqrt{\varepsilon^2 -\Delta^2}}
\ln\left|\frac{(D-\omega_0/2+\sqrt{\varepsilon^2
-\Delta^2})(\omega_0/2+\sqrt{\varepsilon^2
-\Delta^2})}{(D-\omega_0/2-\sqrt{\varepsilon^2
-\Delta^2})(\omega_0/2-\sqrt{\varepsilon^2
-\Delta^2})}\right|\right]

\\

&&+\frac{\Delta}{4\pi\rho v_F^2}i\pi\rm{sgn}(\varepsilon) \{
\theta(\Delta<|\varepsilon|<\sqrt{\frac{\omega_0^2}{4}+\Delta^2})(1-\frac{\frac{\omega_0}{2}
}{\sqrt{\varepsilon^2 -\Delta^2}})\\

&&  -
\theta(\Delta<|\varepsilon|<\sqrt{(D-\frac{\omega_0}{2})^2+\Delta^2})(1+\frac{\frac{\omega_0}{2} }{\sqrt{\varepsilon^2 -\Delta^2}})\}\\

\end{array}\label{app2}
\end{equation}
 where $\rho $ is the graphene planar density, $D$ is  a high-energy cutoff of the graphene bandwidth. When the
 gate voltage is taken into account, the Green's functions for the irradiated graphene are
 obtained by changing $\varepsilon$ to $\varepsilon-V_g$ in
 equations (\ref{app1})-(\ref{app2}).

\newpage

\section*{Supplementary materials}

\section{Rotating wave approximation}
Here we give full details of the rotating wave approximation to the Hamiltonian $H_G$ as defined in the main text, i.e.
\begin{equation}
\begin{array}{cll}
\hspace*{-2cm}H_G&=&H_0+H_I
=
\sum\limits_{\mathbf{k}s\tau}\epsilon_{s\mathbf{k}}a_{s\mathbf{k}\tau}^\dag
a_{s\mathbf{k}\tau}- ev_FA(t)\sum\limits_\tau
\sum\limits_{\mathbf{k}ss'}d_{ss',\mathbf{k}}
a_{s\mathbf{k}\tau}^\dag a_{s'\mathbf{k}\tau}
\\
\end{array}
\end{equation}
\begin{equation}
\hspace*{-2cm}H_0=\sum\limits_{\mathbf{k}\tau}[\epsilon_{+\mathbf{k}}a_{+\mathbf{k}\tau}^\dag
a_{+\mathbf{k}\tau}+\epsilon_{-\mathbf{k}}a_{-\mathbf{k}\tau}^\dag
a_{-\mathbf{k}\tau}]
\end{equation}
\begin{equation}
\hspace*{-2cm} H_I=-ev_FA(t) \sum\limits_{\mathbf{k}\tau}[d_{++,\mathbf{k}}
a_{+\mathbf{k}\tau}^\dag a_{+\mathbf{k}\tau}+d_{--,\mathbf{k}}
a_{-\mathbf{k}\tau}^\dag a_{-\mathbf{k}\tau}+d_{+-,\mathbf{k}}
a_{+\mathbf{k}\tau}^\dag a_{-\mathbf{k}\tau}+d_{-+,\mathbf{k}}
a_{-\mathbf{k}\tau}^\dag a_{+\mathbf{k}\tau}]
\end{equation}
where $A(t)=A_0 e^{i\omega_0 t}+A_0^*e^{-i\omega_0}$. In the
interaction picture, we have
\begin{equation}
\overline{H}=i
\dot{U}U^\dag+UH_GU^\dag=UH_IU^\dag=e^{iH_0t}H_Ie^{-iH_0t}
\end{equation}
Using the formula $$e^A B e^{-A}=\sum\limits_{n=0}^\infty
\frac{1}{n!}C_n,\ \ C_n=[A,C_{n-1}],\ \ C_0=B,$$ we have\\
 $[itH_0,H_I]=$
\begin{equation}
\begin{array}{cll}
&&\hspace*{-3cm}
-\sum\limits_{\mathbf{k}\tau\mathbf{k}'\tau'}itev_FA(t)\{
\epsilon_{+\mathbf{k}}d_{+-,\mathbf{k}'} [a_{+\mathbf{k}\tau}^\dag
a_{+\mathbf{k}\tau}, a_{+\mathbf{k}'\tau'}^\dag
a_{-\mathbf{k}'\tau'}] +\epsilon_{-\mathbf{k}}d_{+-,\mathbf{k}'}
[a_{-\mathbf{k}\tau}^\dag a_{-\mathbf{k}\tau},
a_{+\mathbf{k}'\tau'}^\dag a_{-\mathbf{k}'\tau'}]\\

&&\hspace*{-3cm}+\epsilon_{+\mathbf{k}}d_{-+,\mathbf{k}'}
[a_{+\mathbf{k}\tau}^\dag a_{+\mathbf{k}\tau},
a_{-\mathbf{k}'\tau'}^\dag a_{+\mathbf{k}'\tau'}]
+\epsilon_{-\mathbf{k}}d_{-+,\mathbf{k}'}
[a_{-\mathbf{k}\tau}^\dag a_{-\mathbf{k}\tau},
a_{-\mathbf{k}'\tau'}^\dag a_{+\mathbf{k}'\tau'}]\}\\

&\hspace*{-3cm}=&\hspace*{-3cm}-\sum\limits_{\mathbf{k}\tau\mathbf{k}'\tau'}itev_FA(t)\left\{
\epsilon_{+\mathbf{k}}d_{+-,\mathbf{k}'}
\left(a_{+\mathbf{k}\tau}^\dag[ a_{+\mathbf{k}\tau},
a_{+\mathbf{k}'\tau'}^\dag a_{-\mathbf{k}'\tau'}]
+[a_{+\mathbf{k}\tau}^\dag , a_{+\mathbf{k}'\tau'}^\dag
a_{-\mathbf{k}'\tau'}]a_{+\mathbf{k}\tau}\right)\right.\\

&&\hspace*{-3cm}+\epsilon_{-\mathbf{k}}d_{+-,\mathbf{k}'}\left(
a_{-\mathbf{k}\tau}^\dag[ a_{-\mathbf{k}\tau},
a_{+\mathbf{k}'\tau'}^\dag a_{-\mathbf{k}'\tau'}]
+[a_{-\mathbf{k}\tau}^\dag ,
a_{+\mathbf{k}'\tau'}^\dag a_{-\mathbf{k}'\tau'}]a_{-\mathbf{k}\tau}\right)\\

&&\hspace*{-3cm}+\epsilon_{+\mathbf{k}}d_{-+,\mathbf{k}'}\left(
a_{+\mathbf{k}\tau}^\dag[ a_{+\mathbf{k}\tau},
a_{-\mathbf{k}'\tau'}^\dag a_{+\mathbf{k}'\tau'}]
+[a_{+\mathbf{k}\tau}^\dag , a_{-\mathbf{k}'\tau'}^\dag
a_{+\mathbf{k}'\tau'}]a_{+\mathbf{k}\tau} \right)\\

&&\hspace*{-3cm}+\left.\epsilon_{-\mathbf{k}}d_{-+,\mathbf{k}'}\left(
a_{-\mathbf{k}\tau}^\dag[ a_{-\mathbf{k}\tau},
a_{-\mathbf{k}'\tau'}^\dag
a_{+\mathbf{k}'\tau'}]+[a_{-\mathbf{k}\tau}^\dag ,
a_{-\mathbf{k}'\tau'}^\dag a_{+\mathbf{k}'\tau'}]a_{-\mathbf{k}\tau}\right)\right\}\\
&\hspace*{-3cm}=&\hspace*{-3cm}-\sum\limits_{\mathbf{k}\tau\mathbf{k}'\tau'}itev_FA(t)\left\{
\epsilon_{+\mathbf{k}}d_{+-,\mathbf{k}'}
\delta_{\mathbf{k}\mathbf{k}'}\delta_{\tau\tau'}a_{+\mathbf{k}\tau}^\dag
a_{-\mathbf{k}'\tau'}\right.

-\epsilon_{-\mathbf{k}}d_{+-,\mathbf{k}'}
\delta_{\mathbf{k}\mathbf{k}'}\delta_{\tau\tau'}a_{+\mathbf{k}'\tau'}^\dag
a_{-\mathbf{k}\tau} \\

&&\hspace*{-3cm}-\epsilon_{+\mathbf{k}}d_{-+,\mathbf{k}'}\delta_{\mathbf{k}\mathbf{k}'}\delta_{\tau\tau'}
a_{-\mathbf{k}'\tau'}^\dag a_{+\mathbf{k}\tau}

+\left.\epsilon_{-\mathbf{k}}d_{-+,\mathbf{k}'}
\delta_{\mathbf{k}\mathbf{k}'}\delta_{\tau\tau'}
a_{-\mathbf{k}\tau}^\dag
a_{+\mathbf{k}'\tau'}\right\},\\
\end{array}
\end{equation}
which means that
\begin{equation}
\begin{array}{cll}
&&\hspace*{-2cm}[itH_0,H_I] = -\sum\limits_{\mathbf{k}\tau }itev_FA(t)\left\{
\epsilon_{+\mathbf{k}}d_{+-,\mathbf{k}} a_{+\mathbf{k}\tau}^\dag
a_{-\mathbf{k}\tau}\right.

-\epsilon_{-\mathbf{k}}d_{+-,\mathbf{k}} a_{+\mathbf{k}\tau}^\dag
a_{-\mathbf{k}\tau} \\

&&-\epsilon_{+\mathbf{k}}d_{-+,\mathbf{k}}
a_{-\mathbf{k}\tau}^\dag a_{+\mathbf{k}\tau}

+\left.\epsilon_{-\mathbf{k}}d_{-+,\mathbf{k}}
a_{-\mathbf{k}\tau}^\dag
a_{+\mathbf{k}\tau}\right\}\\

&=&-\sum\limits_{\mathbf{k}\tau }itev_FA(t)\left\{
(\epsilon_{+\mathbf{k}}-\epsilon_{-\mathbf{k}})d_{+-,\mathbf{k}}
a_{+\mathbf{k}\tau}^\dag a_{-\mathbf{k}\tau}

-(\epsilon_{+\mathbf{k}}-\epsilon_{-\mathbf{k}})d_{-+,\mathbf{k}}
a_{-\mathbf{k}\tau}^\dag a_{+\mathbf{k}\tau}\right\},

\end{array}
\end{equation}
thus
\begin{equation}
\begin{array}{cll}
&&\hspace*{-2cm}\overline{H}=H_I-ev_FA(t)\sum\limits_{\mathbf{k}\tau }\left\{
it(\epsilon_{+\mathbf{k}}-\epsilon_{-\mathbf{k}})d_{+-,\mathbf{k}}
a_{+\mathbf{k}\tau}^\dag a_{-\mathbf{k}\tau}

-it(\epsilon_{+\mathbf{k}}-\epsilon_{-\mathbf{k}})d_{-+,\mathbf{k}}
a_{-\mathbf{k}\tau}^\dag a_{+\mathbf{k}\tau}\right\}\\

&&\hspace*{-3cm}-ev_FA(t)\sum\limits_{\mathbf{k}\tau }\left\{
\frac{1}{2!}[it(\epsilon_{+\mathbf{k}}-\epsilon_{-\mathbf{k}})]^2d_{+-,\mathbf{k}}
a_{+\mathbf{k}\tau}^\dag a_{-\mathbf{k}\tau}

+\frac{1}{2!}[-it(\epsilon_{+\mathbf{k}}-\epsilon_{-\mathbf{k}})]^2d_{-+,\mathbf{k}}
a_{-\mathbf{k}\tau}^\dag a_{+\mathbf{k}\tau}\right\}+\cdots\\
&&\hspace*{-3cm}-ev_FA(t)\sum\limits_{\mathbf{k}\tau }\left\{
\frac{1}{n!}[it(\epsilon_{+\mathbf{k}}-\epsilon_{-\mathbf{k}})]^nd_{+-,\mathbf{k}}
a_{+\mathbf{k}\tau}^\dag a_{-\mathbf{k}\tau}

+\frac{1}{n!}[-it(\epsilon_{+\mathbf{k}}-\epsilon_{-\mathbf{k}})]^nd_{-+,\mathbf{k}}
a_{-\mathbf{k}\tau}^\dag a_{+\mathbf{k}\tau}\right\}+\cdots\\

&\hspace*{-2cm}=&\hspace*{-2cm}-ev_FA(t) \sum\limits_{\mathbf{k}\tau}\left(d_{++,\mathbf{k}}
a_{+\mathbf{k}\tau}^\dag a_{+\mathbf{k}\tau}+d_{--,\mathbf{k}}
a_{-\mathbf{k}\tau}^\dag a_{-\mathbf{k}\tau}\right)\\

&&\hspace*{-2cm}-ev_FA(t)
\sum\limits_{\mathbf{k}\tau}\left\{\left\{1+it(\epsilon_{+\mathbf{k}}-\epsilon_{-\mathbf{k}})
+\frac{1}{2!}[it(\epsilon_{+\mathbf{k}}-\epsilon_{-\mathbf{k}})]^2+\cdots\right\}d_{+-,\mathbf{k}}
a_{+\mathbf{k}\tau}^\dag a_{-\mathbf{k}\tau}\right.\\

&&\hspace*{-2cm}+\left.\left\{
1-it(\epsilon_{+\mathbf{k}}-\epsilon_{-\mathbf{k}})
+\frac{1}{2!}[-it(\epsilon_{+\mathbf{k}}-\epsilon_{-\mathbf{k}})]^2+\cdots\right\}d_{-+,\mathbf{k}}
a_{-\mathbf{k}\tau}^\dag a_{+\mathbf{k}\tau}\right\}\\

&\hspace*{-2cm}=&\hspace*{-2cm}-ev_FA(t) \sum\limits_{\mathbf{k}\tau}\left(d_{++,\mathbf{k}}
a_{+\mathbf{k}\tau}^\dag a_{+\mathbf{k}\tau}+d_{--,\mathbf{k}}
a_{-\mathbf{k}\tau}^\dag a_{-\mathbf{k}\tau}\right)\\

&&\hspace*{-2cm}-ev_FA(t)
\sum\limits_{\mathbf{k}\tau}\left\{e^{it(\epsilon_{+\mathbf{k}}-\epsilon_{-\mathbf{k}})}
d_{+-,\mathbf{k}} a_{+\mathbf{k}\tau}^\dag a_{-\mathbf{k}\tau}

+e^{-it(\epsilon_{+\mathbf{k}}-\epsilon_{-\mathbf{k}})}
d_{-+,\mathbf{k}} a_{-\mathbf{k}\tau}^\dag
a_{+\mathbf{k}\tau}\right\}.
\end{array}\label{nonrwa}
\end{equation}
Under the rotating wave approximation,Eq.(\ref{nonrwa}) can be
approximated as
\begin{equation}
\begin{array}{cll}
\hspace*{-2cm}\overline{H}&=& -ev_F
\sum\limits_{\mathbf{k}\tau}\left\{A_0^*e^{it(\epsilon_{+\mathbf{k}}-\epsilon_{-\mathbf{k}}-\omega_0)}
d_{+-,\mathbf{k}} a_{+\mathbf{k}\tau}^\dag a_{-\mathbf{k}\tau}

+A_0e^{-it(\epsilon_{+\mathbf{k}}-\epsilon_{-\mathbf{k}}-\omega_0)}
d_{-+,\mathbf{k}} a_{-\mathbf{k}\tau}^\dag
a_{+\mathbf{k}\tau}\right\}.
\end{array}\label{rwa}
\end{equation}
The next step is to transform the approximate Hamiltonian back to
the Schr\"{o}dinger picture:
\begin{equation}
\begin{array}{cll}
\hspace*{-2cm} H_I&=& e^{-iH_0t}H_Ie^{iH_0t}=-ev_F
\sum\limits_{\mathbf{k}\tau}\left\{A_0^*e^{-i\omega_0 t}
d_{+-,\mathbf{k}} a_{+\mathbf{k}\tau}^\dag a_{-\mathbf{k}\tau}
+A_0e^{i\omega_0 t} d_{-+,\mathbf{k}} a_{-\mathbf{k}\tau}^\dag
a_{+\mathbf{k}\tau}\right\}
\end{array}\label{rwaaa}
\end{equation}

\section{Electric current formula}
The Hamiltonian
$H=H_G+H_L+H_R+H_T$
is determined by
\begin{equation}
\begin{array}{cll}
\hspace*{-2cm}H_G
&=&
\sum\limits_{\mathbf{k}s\tau}\epsilon_{s\mathbf{k}}a_{s\mathbf{k}\tau}^\dag
a_{s\mathbf{k}\tau}- ev_F
\sum\limits_{\mathbf{k}\tau}\left\{A_0^*e^{-i\omega_0 t}
d_{+-,\mathbf{k}} a_{+\mathbf{k}\tau}^\dag a_{-\mathbf{k}\tau}
+A_0e^{i\omega_0 t} d_{-+,\mathbf{k}} a_{-\mathbf{k}\tau}^\dag
a_{+\mathbf{k}\tau}\right\},
\end{array}
\end{equation}
\begin{equation}
\hspace*{-2cm}H_\lambda=\sum\limits_{\mathbf{q}\lambda \tau}
\varepsilon_{\mathbf{q}\lambda \tau} c_{\mathbf{q}\lambda
\tau}^\dag c_{\mathbf{q}\lambda \tau},\ \ \lambda=L,R,
\end{equation}
\begin{equation}
\hspace*{-2cm}H_T=\frac{1}{\sqrt{N}}\sum\limits_{\mathbf{kq}\lambda\tau
s}\left[T_{\mathbf{k}\lambda\mathbf{q}}c_{\mathbf{q}\lambda
\tau}^\dag a_{s\mathbf{k}\tau}+ {\rm H.c.}\right].
\end{equation}
With the unitary transformation
\begin{equation}
U=\exp\left[-i\frac{\omega_0
t}{2}\sum\limits_{\mathbf{k}\tau}(a_{+\mathbf{k}\tau}^\dag
a_{+\mathbf{k}\tau}-a_{-\mathbf{k}\tau}^\dag
a_{-\mathbf{k}\tau})\right]
\end{equation}
one finds
\begin{equation}
\begin{array}{cll}
\hspace*{-2cm}U^{-1}H_IU&=&-ev_F\exp\left[i\frac{\omega_0
t}{2}\sum\limits_{\mathbf{k}\tau}(a_{+\mathbf{k}\tau}^\dag
a_{+\mathbf{k}\tau}-a_{-\mathbf{k}\tau}^\dag
a_{-\mathbf{k}\tau})\right]\\

&&\times  \sum\limits_{\mathbf{k}\tau}\left\{A_0^*e^{-i\omega_0 t}
d_{+-,\mathbf{k}}^y a_{+\mathbf{k}\tau}^\dag a_{-\mathbf{k}\tau}
+A_0e^{i\omega_0 t} d_{-+,\mathbf{k}}
a_{-\mathbf{k}\tau}^\dag a_{+\mathbf{k}\tau}\right\}\\

&&\times \exp\left[-i\frac{\omega_0
t}{2}\sum\limits_{\mathbf{k}\tau}(a_{+\mathbf{k}\tau}^\dag
a_{+\mathbf{k}\tau}-a_{-\mathbf{k}\tau}^\dag
a_{-\mathbf{k}\tau})\right]\\

&=&-ev_F\exp\left[i\frac{\omega_0
t}{2}\sum\limits_{\mathbf{k}\tau}(a_{+\mathbf{k}\tau}^\dag
a_{+\mathbf{k}\tau}-a_{-\mathbf{k}\tau}^\dag
a_{-\mathbf{k}\tau})\right]

\sum\limits_{\mathbf{k}\tau} A_0^*e^{-i\omega_0 t}
d_{+-,\mathbf{k}} a_{+\mathbf{k}\tau}^\dag a_{-\mathbf{k}\tau}\\

&&\times \exp\left[-i\frac{\omega_0
t}{2}\sum\limits_{\mathbf{k}\tau}(a_{+\mathbf{k}\tau}^\dag
a_{+\mathbf{k}\tau}-a_{-\mathbf{k}\tau}^\dag
a_{-\mathbf{k}\tau})\right]\\

 &&+  ev_F\exp\left[i\frac{\omega_0
t}{2}\sum\limits_{\mathbf{k}\tau}(a_{+\mathbf{k}\tau}^\dag
a_{+\mathbf{k}\tau}-a_{-\mathbf{k}\tau}^\dag
a_{-\mathbf{k}\tau})\right]\sum\limits_{\mathbf{k}\tau}
A_0e^{i\omega_0 t} d_{-+,\mathbf{k}}
a_{-\mathbf{k}\tau}^\dag a_{+\mathbf{k}\tau}\\

&&\times \exp\left[-i\frac{\omega_0
t}{2}\sum\limits_{\mathbf{k}\tau}(a_{+\mathbf{k}\tau}^\dag
a_{+\mathbf{k}\tau}-a_{-\mathbf{k}\tau}^\dag
a_{-\mathbf{k}\tau})\right]\\
\end{array}
\end{equation}
The following applies
\begin{equation}
\begin{array}{cll}
 &\hspace*{-3cm}&\hspace*{-3cm}[i\frac{\omega_0
t}{2}\sum\limits_{\mathbf{k}\tau}(a_{+\mathbf{k}\tau}^\dag
a_{+\mathbf{k}\tau}-a_{-\mathbf{k}\tau}^\dag
a_{-\mathbf{k}\tau}),\sum\limits_{\mathbf{k}\tau}
A_0^*e^{-i\omega_0 t} d_{+-,\mathbf{k}}  a_{+\mathbf{k}\tau}^\dag
a_{-\mathbf{k}\tau}]\\

&=&\sum\limits_{\mathbf{k}\tau\mathbf{k}'\tau'}\frac{i\omega_0
t}{2}A_0^*e^{-i\omega_0 t} d_{+-,\mathbf{k}'}
[a_{+\mathbf{k}\tau}^\dag
a_{+\mathbf{k}\tau}-a_{-\mathbf{k}\tau}^\dag
a_{-\mathbf{k}\tau},a_{+\mathbf{k}'\tau'}^\dag
a_{-\mathbf{k}'\tau'}]\\

&=&\sum\limits_{\mathbf{k}\tau\mathbf{k}'\tau'}\frac{i\omega_0
t}{2}A_0^*e^{-i\omega_0 t} d_{+-,\mathbf{k}'}
\{[a_{+\mathbf{k}\tau}^\dag
a_{+\mathbf{k}\tau},a_{+\mathbf{k}'\tau'}^\dag
a_{-\mathbf{k}'\tau'}]

-[a_{-\mathbf{k}\tau}^\dag
a_{-\mathbf{k}\tau},a_{+\mathbf{k}'\tau'}^\dag
a_{-\mathbf{k}'\tau'}]\}\\

&=&\sum\limits_{\mathbf{k}\tau\mathbf{k}'\tau'}\frac{i\omega_0
t}{2}A_0^*e^{-i\omega_0 t} d_{+-,\mathbf{k}'}
\{a_{+\mathbf{k}\tau}^\dag[
a_{+\mathbf{k}\tau},a_{+\mathbf{k}'\tau'}^\dag
a_{-\mathbf{k}'\tau'}]

+[a_{+\mathbf{k}\tau}^\dag ,a_{+\mathbf{k}'\tau'}^\dag
a_{-\mathbf{k}'\tau'}]a_{+\mathbf{k}\tau}\\

&&-a_{-\mathbf{k}\tau}^\dag[
a_{-\mathbf{k}\tau},a_{+\mathbf{k}'\tau'}^\dag
a_{-\mathbf{k}'\tau'}]

-[a_{-\mathbf{k}\tau}^\dag ,a_{+\mathbf{k}'\tau'}^\dag
a_{-\mathbf{k}'\tau'}]a_{-\mathbf{k}\tau}\}\\

&=&\sum\limits_{\mathbf{k}\tau\mathbf{k}'\tau'}\frac{i\omega_0
t}{2}A_0^*e^{-i\omega_0 t} d_{+-,\mathbf{k}'}
\{\delta_{\mathbf{k}\mathbf{k}'}\delta_{\tau\tau'}a_{+\mathbf{k}\tau}^\dag
a_{-\mathbf{k}'\tau'}

 +\delta_{\mathbf{k}\mathbf{k}'}\delta_{\tau\tau'}a_{+\mathbf{k}'\tau'}^\dag
a_{-\mathbf{k}\tau}\}\\

&=&\sum\limits_{\mathbf{k}\tau } i\omega_0 t A_0^*e^{-i\omega_0 t}
d_{+-,\mathbf{k}}  a_{+\mathbf{k}\tau}^\dag a_{-\mathbf{k}\tau}.\\
\end{array}
\end{equation}

Furthermore, one finds

\begin{equation}
\begin{array}{cll}
&&[i\frac{\omega_0
t}{2}\sum\limits_{\mathbf{k}\tau}(a_{+\mathbf{k}\tau}^\dag
a_{+\mathbf{k}\tau}-a_{-\mathbf{k}\tau}^\dag
a_{-\mathbf{k}\tau}),\sum\limits_{\mathbf{k}\tau} A_0 e^{i\omega_0
t} d_{-+,\mathbf{k}}  a_{-\mathbf{k}\tau}^\dag
a_{+\mathbf{k}\tau}]\\

&=&\sum\limits_{\mathbf{k}\tau\mathbf{k}'\tau'}\frac{i\omega_0
t}{2}A_0e^{i\omega_0 t} d_{-+,\mathbf{k}'}
[a_{+\mathbf{k}\tau}^\dag
a_{+\mathbf{k}\tau}-a_{-\mathbf{k}\tau}^\dag
a_{-\mathbf{k}\tau},a_{-\mathbf{k}'\tau'}^\dag
a_{+\mathbf{k}'\tau'}]\\

&=&\sum\limits_{\mathbf{k}\tau\mathbf{k}'\tau'}\frac{i\omega_0
t}{2}A_0e^{i\omega_0 t} d_{-+,\mathbf{k}'}
\{[a_{+\mathbf{k}\tau}^\dag
a_{+\mathbf{k}\tau},a_{-\mathbf{k}'\tau'}^\dag
a_{+\mathbf{k}'\tau'}]

-[a_{-\mathbf{k}\tau}^\dag
a_{-\mathbf{k}\tau},a_{-\mathbf{k}'\tau'}^\dag
a_{+\mathbf{k}'\tau'}]\}\\

&=&\sum\limits_{\mathbf{k}\tau\mathbf{k}'\tau'}\frac{i\omega_0
t}{2}A_0e^{i\omega_0 t} d_{-+,\mathbf{k}'}
\{a_{+\mathbf{k}\tau}^\dag[
a_{+\mathbf{k}\tau},a_{-\mathbf{k}'\tau'}^\dag
a_{+\mathbf{k}'\tau'}]

+[a_{+\mathbf{k}\tau}^\dag ,a_{-\mathbf{k}'\tau'}^\dag
a_{+\mathbf{k}'\tau'}]a_{+\mathbf{k}\tau}\\

&&-a_{-\mathbf{k}\tau}^\dag[
a_{-\mathbf{k}\tau},a_{-\mathbf{k}'\tau'}^\dag
a_{+\mathbf{k}'\tau'}]

-[a_{-\mathbf{k}\tau}^\dag ,a_{-\mathbf{k}'\tau'}^\dag
a_{+\mathbf{k}'\tau'}]a_{-\mathbf{k}\tau}\}\\

&=&\sum\limits_{\mathbf{k}\tau\mathbf{k}'\tau'}\frac{i\omega_0
t}{2}A_0e^{i\omega_0 t} d_{-+,\mathbf{k}'}
\{-\delta_{\mathbf{k}\mathbf{k}'}\delta_{\tau\tau'}a_{-\mathbf{k}'\tau'}^\dag
a_{+\mathbf{k}'\tau'}

 -\delta_{\mathbf{k}\mathbf{k}'}\delta_{\tau\tau'}a_{-\mathbf{k}\tau}^\dag
a_{+\mathbf{k}'\tau'}\}\\

&=&\sum\limits_{\mathbf{k}\tau } (-i\omega_0 t) A_0e^{i\omega_0 t}
d_{-+,\mathbf{k}}  a_{-\mathbf{k}\tau}^\dag a_{+\mathbf{k}\tau}.
\end{array}
\end{equation}
Another straightforward calculation is that
\begin{equation}
\begin{array}{cll}
U^{-1}H_GU &=&
\sum\limits_{\mathbf{k}s\tau}\epsilon_{s\mathbf{k}}a_{s\mathbf{k}\tau}^\dag
a_{s\mathbf{k}\tau}- ev_F \sum\limits_{\mathbf{k}\tau}\left\{A_0^*
d_{+-,\mathbf{k}} a_{+\mathbf{k}\tau}^\dag a_{-\mathbf{k}\tau}

+A_0 d_{-+,\mathbf{k}} a_{-\mathbf{k}\tau}^\dag
a_{+\mathbf{k}\tau}\right\}

\end{array}
\end{equation}
\begin{equation}
\begin{array}{cll}
U^{-1}H_TU&=&\frac{1}{\sqrt{N}}\sum\limits_{\mathbf{kq}\lambda\tau}\left[T_{\mathbf{k}\lambda\mathbf{q}}e^{-i\frac{\omega_0
t}{2}}c_{\mathbf{q}\lambda \tau}^\dag
a_{+\mathbf{k}\tau}+T_{\mathbf{k}\lambda\mathbf{q}}e^{i\frac{\omega_0
t}{2}}c_{\mathbf{q}\lambda \tau}^\dag a_{-\mathbf{k}\tau}+
{\rm H.c.}\right]\\

& =&\frac{1}{\sqrt{N}}\sum\limits_{\mathbf{kq}\lambda
s\tau}\left[T_{\mathbf{k}\lambda
s\mathbf{q}}(t)c_{\mathbf{q}\lambda \tau}^\dag
a_{s\mathbf{k}\tau}+ {\rm H.c.}\right]
\end{array}
\end{equation}
where $T_{\mathbf{k}\lambda s\mathbf{q}}(t)=T_{\mathbf{k}\lambda
\mathbf{q}}e^\frac{-is\omega_0 t}{2}$. The final Hamiltonian is
expressed in the rotating reference as
\begin{equation}
\begin{array}{cll}
\widetilde{H}&=&U^{-1}HU+i\frac{d U^{-1}}{dt}U\\

&=&\sum\limits_{\mathbf{k}\tau}\left[(\epsilon_{+\mathbf{k}}-\frac{\omega_0}{2})a_{+\mathbf{k}\tau}^\dag
a_{+\mathbf{k}\tau}+(\epsilon_{-\mathbf{k}}+\frac{\omega_0}{2})a_{-\mathbf{k}\tau}^\dag
a_{-\mathbf{k}\tau}\right]\\
&&- ev_F \sum\limits_{\mathbf{k}\tau}\left\{A_0^*
d_{+-,\mathbf{k}} a_{+\mathbf{k}\tau}^\dag a_{-\mathbf{k}\tau}

+A_0 d_{-+,\mathbf{k}}  a_{-\mathbf{k}\tau}^\dag
a_{+\mathbf{k}\tau}\right\}\\

&&+\sum\limits_{\mathbf{q}\lambda \tau}
\varepsilon_{\mathbf{q}\lambda \tau} c_{\mathbf{q}\lambda
\tau}^\dag c_{\mathbf{q}\lambda
\tau}+\frac{1}{\sqrt{N}}\sum\limits_{\mathbf{kq}\lambda
s\tau}\left[T_{\mathbf{k}\lambda
s\mathbf{q}}(t)c_{\mathbf{q}\lambda \tau}^\dag
a_{s\mathbf{k}\tau}+ {\rm H.c.}\right].\\

\end{array}
\end{equation}
%

%
Assuming  that
$p_x>>p_y$, then  $\phi\approx 0,\ \ \cos\phi=1$ and
 the Hamiltonian is further simplified to
\begin{equation}
\begin{array}{cll}
\widetilde{H}&=&\widetilde{H}_G+\widetilde{H}_L+\widetilde{H}_R+\widetilde{H}_T\\

&=&\sum\limits_{\mathbf{k}s\tau}\widetilde{\epsilon}_{s\mathbf{k}}a_{s\mathbf{k}\tau}^\dag
a_{s\mathbf{k}\tau} +  \Delta\sum\limits_{\mathbf{k}\tau}\left(
a_{+\mathbf{k}\tau}^\dag a_{-\mathbf{k}\tau}

+ a_{-\mathbf{k}\tau}^\dag
a_{+\mathbf{k}\tau}\right)\\

&&+\sum\limits_{\mathbf{q}\lambda \tau}
\varepsilon_{\mathbf{q}\lambda \tau} c_{\mathbf{q}\lambda
\tau}^\dag c_{\mathbf{q}\lambda
\tau}+\frac{1}{\sqrt{N}}\sum\limits_{\mathbf{kq}\lambda
s\tau}\left[T_{\mathbf{k}\lambda
s\mathbf{q}}(t)c_{\mathbf{q}\lambda \tau}^\dag
a_{s\mathbf{k}\tau}+ {\rm H.c.}\right],\\

\end{array}\label{rothamil}
\end{equation}
where
$\widetilde{\epsilon}_{s\mathbf{k}}=\epsilon_{s\mathbf{k}}-\frac{s\omega_0}{2}$
and $\Delta=\frac{ev_FE_0}{2\omega_0}$. The current can be
calculated from the time variation of the occupation number operator
of the left electrode.
\begin{equation}
\hspace*{-2.5cm}I_L=e\langle\dot{ \mathcal{N}_L}\rangle=\frac{ie}{\hbar}\langle
[\widetilde{H},\mathcal{N}_L]\rangle
=\frac{ie}{\hbar}\sum\limits_{\mathbf{q}\tau}\langle
[H,c_{\mathbf{q}L\tau}^\dag
c_{\mathbf{q}L\tau}]\rangle=\frac{ie}{\hbar}\sum\limits_{\mathbf{q}\tau}[\langle
[\widetilde{H}_T,c_{\mathbf{q}L\tau}^\dag]
c_{\mathbf{q}L\tau}\rangle+\langle c_{\mathbf{q}L\tau}^\dag
[\widetilde{H}_T,c_{\mathbf{q}L\tau}]\rangle], \label{jl0}
\end{equation}
The commutation relation reads
\begin{equation}
\begin{array}{cll}
[\widetilde{H}_T,c_{\mathbf{q}L\tau}^\dag]&=&\frac{1}{\sqrt{N}}\sum\limits_{\mathbf{k}\mathbf{q}'s\tau'}
T_{\mathbf{k}Ls\mathbf{q}}^*(t)[a_{s\mathbf{k}\tau'}^\dag
c_{\mathbf{q}'L\tau'},c_{\mathbf{q}L\tau}^\dag]\\
&=&\frac{1}{\sqrt{N}}\sum\limits_{\mathbf{k}\mathbf{q}'s\tau'}
T_{\mathbf{k}Ls\mathbf{q}'}^*(t) a_{s\mathbf{k}\tau'}^\dag
\{c_{\mathbf{q}'L\tau'},c_{\mathbf{q}L\tau}^\dag\}
=\frac{1}{\sqrt{N}}\sum\limits_{\mathbf{k}s}
T_{\mathbf{k}Ls\mathbf{q}}^*(t)a_{s\mathbf{k}\tau}^\dag
\end{array}
\end{equation}
\begin{equation}
[\widetilde{H}_T,c_{\mathbf{q}L\tau}]=-\frac{1}{\sqrt{N}}\sum\limits_{\mathbf{k}s}
T_{\mathbf{k}Ls\mathbf{q}}(t)a_{s\mathbf{k}\tau}.
\end{equation}
Thus
\begin{equation}
I_L=e\langle\dot{
\mathcal{N}}\rangle=\frac{ie}{\hbar}\frac{1}{\sqrt{N}}\sum\limits_{\mathbf{k}\mathbf{q}s\tau}[T_{\mathbf{k}Ls\mathbf{q}}^*(t)\langle
a_{s\mathbf{k}\tau}^\dag
c_{\mathbf{q}L\tau}\rangle-T_{\mathbf{k}Ls\mathbf{q}}(t)\langle
c_{\mathbf{q}L\tau}^\dag a_{s\mathbf{k}\tau}\rangle].
\end{equation}
Define $G_{\mathbf{k}s,\mathbf{q}L}^{\tau'\tau,<}(t,t')=i\langle
c_{\mathbf{q}L\tau}^\dag(t') a_{s\mathbf{k}\tau'}(t) \rangle$,
then
\begin{equation}
I_L=-\frac{2e}{\hbar} {\rm Re}\frac{1}{\sqrt{N}}\sum\limits_{\mathbf{k}\mathbf{q}s\tau}T_{\mathbf{k}Ls\mathbf{q}}(t)
G_{\mathbf{k}s,\mathbf{q}L}^{\tau\tau,<}(t,t) .\label{jl}
\end{equation}

To obtain
$G_{\mathbf{k}s,\mathbf{q}L}^{\tau'\tau,<}(t,t')=i\langle
c_{\mathbf{q}L\tau}^\dag(t') a_{s\mathbf{k}\tau'}(t) \rangle$, one
needs to find the time-ordered Green's function, defined as
$G_{\mathbf{k}s,\mathbf{q}L}^{\tau\tau',t}(t,t')=-i\langle
T\{a_{s\mathbf{k}\tau}(t)c_{\mathbf{q}L\tau'}^\dag(t') \}\rangle$.
Using the equation-of-motion method we write
\begin{equation}
\begin{array}{cll}
\hspace*{-2.5cm}&&\hspace*{-2.5cm}-i\frac{\partial}{\partial
t'}G_{\mathbf{k}s,\mathbf{q}L}^{\tau\tau',t}(t,t')\\
 &\hspace*{-2.5cm}=&\hspace*{-2.5cm}-i\langle
T\{a_{s\mathbf{k}\tau}(t')[\widetilde{H},c_{\mathbf{q}L\tau'}^\dag(t')]
\}\rangle\\
&\hspace*{-2.5cm}=&\hspace*{-2.5cm}\varepsilon_{\mathbf{q}L\tau'}[-i\langle
T\{a_{s\mathbf{k}\tau}(t)c_{\mathbf{q}L\tau'}^\dag(t') \}\rangle]
 +\frac{1}{\sqrt{N}}\sum\limits_{\mathbf{k}'s'} T_{\mathbf{k}'Ls'\mathbf{q}}^*(t')\{
[-i\langle
T\{a_{s\mathbf{k}\tau}(t)a_{s'\mathbf{k}'\tau'}^\dag(t')
\}\rangle]\}\\
&\hspace*{-2.5cm}=&\hspace*{-2.5cm}\varepsilon_{\mathbf{q}L\tau'}G_{\mathbf{k}s,\mathbf{q}L}^{\tau\tau',t}(t,t')+\frac{1}{\sqrt{N}}\sum\limits_{\mathbf{k}'s'}
T_{\mathbf{k}'Ls'\mathbf{q}}^*(t')G_{\mathbf{k}s,\mathbf{k}'s'}^{\tau\tau',t}(t,t'),
\end{array}\label{moe1}
\end{equation}
where $G_{\mathbf{k}s,\mathbf{k}'s'}^{\tau\tau',t}(t,t')=-i\langle
T\{a_{s\mathbf{k}\tau}(t)a_{s'\mathbf{k}'\tau'}^\dag(t')
\}\rangle$. Eq.(\ref{moe1}) can be expressed as
\begin{equation}
\begin{array}{cll}
G_{\mathbf{k}s,\mathbf{q}L}^{\tau\tau',t}(t,t')

&=&\frac{1}{\sqrt{N}}\sum\limits_{\mathbf{k}'s'} \int dt_1
G_{\mathbf{k}s,\mathbf{k}'s'}^{\tau\tau',t}(t,t_1)T_{\mathbf{k}'Ls'\mathbf{q}}^*(t_1)g_{\mathbf{q}L}^{\tau'\tau',t}(t_1-t')
\end{array}\label{moe2}
\end{equation}
Make the analytic continuation to Eq.(\ref{moe2}),
\begin{equation}
\begin{array}{cll}
\hspace*{-3cm}G_{\mathbf{k}s,\mathbf{q}L}^{\tau\tau',<}(t,t')

&=&\frac{1}{\sqrt{N}}\sum\limits_{\mathbf{k}'s'} \int
dt_1T_{\mathbf{k}'Ls'\mathbf{q}}^*(t_1)
[G_{\mathbf{k}s,\mathbf{k}'s'}^{\tau\tau',r}(t,t_1)g_{\mathbf{q}L}^{\tau'\tau',<}(t_1-t')
+G_{\mathbf{k}s,\mathbf{k}'s'}^{\tau\tau',<}(t,t_1)g_{\mathbf{q}L}^{\tau'\tau',a}(t_1-t')]
\end{array}\label{analy}
\end{equation}
Substitute Eq.(\ref{analy}) in Eq.(\ref{jl}),
\begin{equation}
\begin{array}{cll}
\hspace*{-4.3cm}I_L&\hspace*{-2.2cm}=&\hspace*{-2cm}-\frac{2e}{\hbar}{\rm Re}\frac{1}{N}\sum\limits_{\mathbf{k}\mathbf{q}s\tau}
\sum\limits_{\mathbf{k}'s'} \int
dt_1T_{\mathbf{k}Ls\mathbf{q}}(t)T_{\mathbf{k}'Ls'\mathbf{q}}^*(t_1)
[G_{\mathbf{k}s,\mathbf{k}'s'}^{\tau\tau,r}(t,t_1)g_{\mathbf{q}L}^{\tau\tau,<}(t_1-t)  \nonumber \\
&\hspace*{-2cm}+& \hspace*{-2cm} G_{\mathbf{k}s,\mathbf{k}'s'}^{\tau\tau,<}(t,t_1)g_{\mathbf{q}L}^{\tau\tau,a}(t_1-t)] \nonumber \\
&\hspace*{-2cm}=&\hspace*{-2cm}-\frac{2e}{\hbar}{\rm Re}\frac{1}{N}\sum\limits_{\mathbf{k}\mathbf{q}s\tau}
\sum\limits_{\mathbf{k}'s'} \int\frac{d\varepsilon}{2\pi}\int dt_1
e^{-i\varepsilon(t_1-t)}T_{\mathbf{k}Ls\mathbf{q}}(t)T_{\mathbf{k}'Ls'\mathbf{q}}^*(t_1)
[G_{\mathbf{k}s,\mathbf{k}'s'}^{\tau\tau,r}(t,t_1)g_{\mathbf{q}L}^{\tau\tau,<}(\varepsilon)  \nonumber\\
&\hspace*{-2cm} +& \hspace*{-2cm}G_{\mathbf{k}s,\mathbf{k}'s'}^{\tau\tau,<}(t,t_1)g_{\mathbf{q}L}^{\tau\tau,a}(\varepsilon)]  \nonumber\\
&\hspace*{-2cm}=&\hspace*{-2cm}-\frac{2e}{\hbar}{\rm Re}\frac{1}{N}\sum\limits_{\mathbf{k}\mathbf{q}s\tau}
\sum\limits_{\mathbf{k}'s'}\int \frac{d\varepsilon}{2\pi}\int dt_1
e^{-i\varepsilon(t_1-t)}T_{\mathbf{k}L\mathbf{q}}T_{\mathbf{k}'L\mathbf{q}}^*e^{-\frac{is\omega_0
t}{2}}e^{\frac{is'\omega_0 t_1}{2}}  \nonumber\\
&&\hspace*{-2cm}\times  [G_{\mathbf{k}s,\mathbf{k}'s'}^{\tau\tau,r}(t,t_1)g_{\mathbf{q}L}^{\tau\tau,<}(\varepsilon)
+G_{\mathbf{k}s,\mathbf{k}'s'}^{\tau\tau,<}(t,t_1)g_{\mathbf{q}L}^{\tau\tau,a}(\varepsilon)]   \nonumber\\
\end{array}
\label{j2}
\end{equation}
where $g_{\mathbf{q}\lambda}^{\tau\tau,r,a}(\varepsilon)
=\frac{1}{\varepsilon-\varepsilon_{\mathbf{q}\lambda\tau}\pm
i\eta}$ and $g_{\mathbf{q}\lambda}^{\tau\tau,<}(\varepsilon)=i2\pi
f_\lambda(\varepsilon_{\mathbf{q}\lambda\tau})\delta(\varepsilon-\varepsilon_{\mathbf{q}\lambda\tau})$
with $f_\lambda(\varepsilon_{\mathbf{q}\lambda\tau})$ denoting the
Fermi distribution function in the $\lambda$ ferromagnetic
electrode. Let
$\Gamma_{L\mathbf{k}\mathbf{k}'}^\tau(\varepsilon)=2\pi\sum\limits_\mathbf{q}
T_{\mathbf{k}L\mathbf{q}}^*T_{\mathbf{k}'L\mathbf{q}}
\delta(\varepsilon-\varepsilon_{\mathbf{q}L\tau})$, then
Eq.(\ref{j2}) can be further expressed as
\begin{equation}
\begin{array}{cll}
I_L

&=&-\frac{e}{\hbar}Re\frac{1}{N}\sum\limits_{\mathbf{k}s\tau}
\sum\limits_{\mathbf{k}'s'}\int \frac{d\varepsilon}{2\pi}\int dt_1
e^{-i\varepsilon(t_1-t)}e^{-\frac{is\omega_0
t}{2}}e^{\frac{is'\omega_0 t_1}{2}}\\
&&\times\{2G_{\mathbf{k}s,\mathbf{k}'s'}^{\tau\tau,r}(t,t_1)
f_{L}(\varepsilon)i\Gamma_{L,\mathbf{k}'\mathbf{k}}^{\tau}
+G_{\mathbf{k}s,\mathbf{k}'s'}^{\tau\tau,<}(t,t_1)i\Gamma_{L,\mathbf{k}'\mathbf{k}}^{\tau}\}.\\
\end{array} \label{j20}
\end{equation}

\section{Derivation of Green's functions}
\subsection{The Green's function for the central region coupled to the electrodes}
We note the following commutation relations
\begin{equation}
\begin{array}{cll}
[a_{s\mathbf{k}\tau},\widetilde{H}]&=&[a_{s\mathbf{k}\tau},\widetilde{H}_G]+[a_{s\mathbf{k}\tau},\widetilde{H}_T]
=\widetilde{\epsilon}_{s\mathbf{k}}a_{s\mathbf{k}\tau}+\delta_{s+}\Delta
a_{-\mathbf{k}\tau}+\delta_{s-}\Delta a_{+\mathbf{k}\tau}\\
&&+\frac{1}{\sqrt{N}}\sum\limits_{\mathbf{q}\lambda} T_{\mathbf{k}\lambda
s\mathbf{q}}^*(t) c_{\mathbf{q}\lambda\tau},
\end{array}
\end{equation}
\begin{equation}
[c_{\mathbf{q}\lambda\tau},\widetilde{H}]=[c_{\mathbf{q}\lambda\tau},\widetilde{H}_\lambda]
+[c_{\mathbf{q}\lambda\tau},\widetilde{H}_T]=\varepsilon_{\mathbf{q}\lambda
\tau}c_{\mathbf{q}\lambda\tau}+\frac{1}{\sqrt{N}}\sum\limits_{\mathbf{k}
s}T_{\mathbf{k}\lambda s\mathbf{q}}(t) a_{s\mathbf{k}\tau},
\end{equation}
with
$\widetilde{\epsilon}_{s\mathbf{k}}=\epsilon_{s\mathbf{k}}-s\omega_0/2=sv_F|\mathbf{k}|-s\omega_0/2$.
The retarded Green's function
$G_{\mathbf{k}s,\mathbf{k}'s'}^{\tau\tau',r}(t,t')=-i\theta(t-t')\langle
\{a_{s\mathbf{k}\tau}(t),a_{s'\mathbf{k}'\tau'}^\dag(t')\}\rangle$  satisfies
\begin{equation}
\begin{array}{cll}
\hspace*{-2cm}i\frac{\partial}{\partial
t}G_{\mathbf{k}s,\mathbf{k}'s'}^{\tau\tau',r}(t,t')
&=&\delta(t-t')\delta_{\mathbf{k}\mathbf{k}'}\delta_{ss'}\delta_{\tau\tau'}-i\theta(t-t')\langle
\{[a_{s\mathbf{k}\tau}(t),\widetilde{H}],a_{s'\mathbf{k}'\tau'}^\dag(t')\}\rangle\\
& \hspace*{-2cm}=&\hspace*{-2cm}\delta(t-t')\delta_{\mathbf{k}\mathbf{k}'}\delta_{ss'}\delta_{\tau\tau'}
+\widetilde{\epsilon}_{s\mathbf{k}}G_{\mathbf{k}s,\mathbf{k}'s'}^{\tau\tau',r}(t,t')
+\delta_{s+}\Delta
G_{\mathbf{k}-,\mathbf{k}'s'}^{\tau\tau',r}(t,t')+\delta_{s-}\Delta
 G_{\mathbf{k}+,\mathbf{k}'s'}^{\tau\tau',r}(t,t')\\
&\hspace*{-2cm}&\hspace*{-2cm}+\frac{1}{\sqrt{N}}\sum\limits_{\mathbf{q}\lambda} T_{\mathbf{k}\lambda
s\mathbf{q}}^*(t)
G_{\mathbf{q}\lambda,\mathbf{k}'s'}^{\tau\tau',r}(t,t').
\end{array}\label{green1}
\end{equation}
In addition,
\begin{equation}
\begin{array}{cll}
i\frac{\partial}{\partial
t}G_{\mathbf{q}\lambda,\mathbf{k}'s'}^{\tau\tau',r}(t,t')
&=&\varepsilon_{\mathbf{q}\lambda
\tau}G_{\mathbf{q}\lambda,\mathbf{k}'s'}^{\tau\tau',r}(t,t')
+\frac{1}{\sqrt{N}}\sum\limits_{\mathbf{k}_1 s_1}T_{\mathbf{k}_1\lambda
s_1\mathbf{q}}(t)
G_{\mathbf{k}_1s_1,\mathbf{k}'s'}^{\tau\tau',r}(t,t'),
\end{array}
\end{equation}
or
\begin{equation}
\begin{array}{cll}
 G_{\mathbf{q}\lambda,\mathbf{k}'s'}^{\tau\tau',r}(t,t')
&=& \frac{1}{\sqrt{N}}\sum\limits_{\mathbf{k}_1 s_1}\int dt_1
g_{\mathbf{q}\lambda}^{\tau\tau,r}(t-t_1)T_{\mathbf{k}_1\lambda
s_1\mathbf{q}}(t_1)
G_{\mathbf{k}_1s_1,\mathbf{k}'s'}^{\tau\tau',r}(t_1,t').
\end{array}\label{green2}
\end{equation}
Substituting Eq.(\ref{green2}) into Eq.(\ref{green1}) leads to
\begin{equation}
\begin{array}{cll}
\hspace*{-4cm}&&\hspace*{-3cm}i\frac{\partial}{\partial
t}G_{\mathbf{k}s,\mathbf{k}'s'}^{\tau\tau',r}(t,t')\\
&\hspace*{-2cm}=&\hspace*{-2cm}\delta(t-t')\delta_{\mathbf{k}\mathbf{k}'}\delta_{ss'}\delta_{\tau\tau'}
+\widetilde{\epsilon}_{s\mathbf{k}}G_{\mathbf{k}s,\mathbf{k}'s'}^{\tau\tau',r}(t,t')
+\delta_{s+}\Delta
G_{\mathbf{k}-,\mathbf{k}'s'}^{\tau\tau',r}(t,t')+\delta_{s-}\Delta G_{\mathbf{k}+,\mathbf{k}'s'}^{\tau\tau',r}(t,t')\\
&&\hspace*{-2cm}+\frac{1}{N}\sum\limits_{\mathbf{q}\lambda}
 \sum\limits_{\mathbf{k}_1 s_1}\int dt_1
T_{\mathbf{k}\lambda
s\mathbf{q}}^*(t)g_{\mathbf{q}\lambda}^{\tau\tau,r}(t-t_1)T_{\mathbf{k}_1\lambda
s_1\mathbf{q}}(t_1)
G_{\mathbf{k}_1s_1,\mathbf{k}'s'}^{\tau\tau',r}(t_1,t')\\
&\hspace*{-2cm}=&\hspace*{-2cm}\delta(t-t')\delta_{\mathbf{k}\mathbf{k}'}\delta_{ss'}\delta_{\tau\tau'}
+\widetilde{\epsilon}_{s\mathbf{k}}G_{\mathbf{k}s,\mathbf{k}'s'}^{\tau\tau',r}(t,t')
+\delta_{s+}\Delta
G_{\mathbf{k}-,\mathbf{k}'s'}^{\tau\tau',r}(t,t')+\delta_{s-}\Delta
 G_{\mathbf{k}+,\mathbf{k}'s'}^{\tau\tau',r}(t,t')\\
&&\hspace*{-2cm}+\frac{1}{N}\sum\limits_{\mathbf{q}\lambda}
 \sum\limits_{\mathbf{k}_1 s_1}\int\frac{d\varepsilon}{2\pi}\int
 dt_1 e^{-i\varepsilon(t-t_1)}
T_{\mathbf{k}\lambda
\mathbf{q}}^*g_{\mathbf{q}\lambda}^{\tau\tau,r}(\varepsilon)T_{\mathbf{k}_1\lambda
\mathbf{q}}e^{\frac{is\omega_0 t}{2}}e^{-\frac{is_1\omega_0
t_1}{2}}
G_{\mathbf{k}_1s_1,\mathbf{k}'s'}^{\tau\tau',r}(t_1,t')\\
&\hspace*{-2cm}=&\hspace*{-2cm}\delta(t-t')\delta_{\mathbf{k}\mathbf{k}'}\delta_{ss'}\delta_{\tau\tau'}
+\widetilde{\epsilon}_{s\mathbf{k}}G_{\mathbf{k}s,\mathbf{k}'s'}^{\tau\tau',r}(t,t')
+\delta_{s+}\Delta
G_{\mathbf{k}-,\mathbf{k}'s'}^{\tau\tau',r}(t,t')+\delta_{s-}\Delta G_{\mathbf{k}+,\mathbf{k}'s'}^{\tau\tau',r}(t,t')\\
\end{array}\label{green3}
\end{equation}
The above expression can be written as
\begin{equation}
\begin{array}{cll}
\hspace*{-2cm}G_{\mathbf{k},\mathbf{k}'}^{\tau\tau',r}(t,t')
=\delta_{\mathbf{k}\mathbf{k}'}\delta_{\tau\tau'}g_\mathbf{k}^{\tau\tau,r}(t-t')
+
 \frac{1}{N}\sum\limits_{\mathbf{k}_1 }\int dt_1dt_2 g_\mathbf{k}^{\tau\tau,r}(t-t_1)\Sigma^{\tau,r}(t_1,t_2)
G_{\mathbf{k}_1,\mathbf{k}'}^{\tau\tau',r}(t_2,t')\\
\end{array}\label{grematn}
\end{equation}
where the quantities without the $s$ and $s'$ indexes means matrixes in $s,s'$ space implicitly, and
\begin{equation}
%
\end{equation}
is the Green's function for graphene without the coupling between the
electrodes and the graphene. The  selfenergy reads
\begin{equation}
\Sigma^{\tau,r}(t,t')=\int\frac{d\varepsilon}{2\pi}\Sigma^{\tau,r}e^{-i\varepsilon(t-t')}\left(\begin{array}{cc}
e^{i\frac{\omega_0}{2}(t-t')}
& e^{i\frac{\omega_0}{2}(t+t') }\\
 e^{-i\frac{\omega_0}{2}(t+t') }&e^{-i\frac{\omega_0}{2}(t-t')}
\end{array}\right)
\end{equation}
with $\Sigma^{\tau,r}=-\frac{i}{2}(\Gamma_L^\tau+\Gamma_R^\tau)$.
Let
$\mathcal{G}^{\tau\tau',r}(t,t')=\frac{1}{N}\sum\limits_{\mathbf{k}\mathbf{k}'
} G_{\mathbf{k},\mathbf{k}'}^{\tau\tau',r}(t,t')$, $ g
^{\tau\tau',r}(t)=\frac{1}{N}\sum\limits_\mathbf{k}g_\mathbf{k}^{\tau\tau',r}(t)$,
Eq.(\ref{grematn}) leads to
\begin{equation}
\begin{array}{cll}
\mathcal{G}^{\tau\tau',r}(t,t')
&=&\delta_{\tau\tau'} g^{\tau\tau,r}(t-t') +
 \int dt_1\int dt_2
 g^{\tau\tau,r}(t-t_1)\Sigma^{\tau,r}(t_1,t_2)
\mathcal{G}^{\tau\tau',r}(t_2,t').\\
\end{array}\label{gremat}
\end{equation}

\subsection{The Green's functions for the central region without coupling to the electrodes}
Below, we attempt to calculate $g^{\tau\tau,r}(t)$.
\begin{equation}
\begin{array}{cll}
&&\hspace*{-3cm} g_{++}^{\tau\tau,r}
= \frac{1}{N}\sum\limits_{\mathbf{k}}
\frac{\varepsilon-\widetilde{\epsilon}_{-\mathbf{k}}}
{(\varepsilon-\widetilde{\epsilon}_{+\mathbf{k}})
(\varepsilon-\widetilde{\epsilon}_{-\mathbf{k}})-\Delta^2}\\
&\hspace*{-2cm}=&\hspace*{-2cm}\frac{S}{(2\pi)^2N}\int
kdkd\phi\frac{\varepsilon-V_g+(v_Fk-\frac{\omega_0}{2})}
{(\varepsilon-V_g)^2-(v_Fk-\frac{\omega_0}{2})^2
-\Delta^2}\\
&\hspace*{-2cm}=&\hspace*{-2cm}\frac{1}{(2\pi)^2\rho}\frac{2\pi}{v_F^2}\int_{-\frac{\omega_0}{2}}^{D-\omega_0/2}
d\varepsilon_k
\frac{(\varepsilon_k+\frac{\omega_0}{2})\left[(\varepsilon-V_g)+\varepsilon_k\right]}
{(\varepsilon-V_g)^2-\varepsilon_k^2 -\Delta^2+i {\rm sgn}(\varepsilon-V_g)\eta}\\
&\hspace*{-2cm}=&\hspace*{-2cm}\frac{1}{(2\pi)^2\rho}\frac{2\pi}{v_F^2}\int_{-\frac{\omega_0}{2}}^{D-\omega_0/2}
d\varepsilon_k
\left[\frac{\varepsilon_k(\varepsilon-V_g+\varepsilon_k)}
{(\varepsilon-V_g)^2-\varepsilon_k^2
-\Delta^2+i{\rm sgn}(\varepsilon-V_g)\eta}
+\frac{\frac{\omega_0}{2}(\varepsilon-V_g+\varepsilon_k)}
{(\varepsilon-V_g)^2-\varepsilon_k^2 -\Delta^2+i{\rm sgn}(\varepsilon-V_g)\eta}\right]\\
&\hspace*{-2cm}=&\hspace*{-2cm}-\frac{1}{(2\pi)^2\rho}\frac{2\pi}{v_F^2}\int_{-\frac{\omega_0}{2}}^{D-\omega_0/2}
d\varepsilon_k
\left[\frac{\varepsilon_k(\varepsilon-V_g+\varepsilon_k)}
{\varepsilon_k^2-(\varepsilon-V_g)^2
+\Delta^2-i{\rm sgn}(\varepsilon-V_g)\eta}+\frac{\frac{\omega_0}{2}(\varepsilon-V_g+\varepsilon_k)}
{\varepsilon_k^2-(\varepsilon-V_g)^2 +\Delta^2-i{\rm sgn}(\varepsilon-V_g)\eta}\right],\\
\end{array}
\end{equation}
If $ |\varepsilon-V_g|<\Delta$, then
\begin{equation}
\begin{array}{cll}
&&-\frac{1}{2\pi\rho
v_F^2}\int_{-\frac{\omega_0}{2}}^{D-\omega_0/2} d\varepsilon_k
\frac{\varepsilon_k(\varepsilon-V_g+\varepsilon_k)}
{\varepsilon_k^2-(\varepsilon-V_g)^2 +\Delta^2 }\\
&=&-\frac{1}{2\pi\rho
v_F^2}\int_{-\frac{\omega_0}{2}}^{D-\omega_0/2} d\varepsilon_k
\left[1+\frac{\varepsilon_k(\varepsilon-V_g)+(\varepsilon-V_g)^2
-\Delta^2}
{\varepsilon_k^2-(\varepsilon-V_g)^2 +\Delta^2 }\right]\\
&=&-\frac{1}{2\pi\rho
v_F^2}\int_{-\frac{\omega_0}{2}}^{D-\omega_0/2} d\varepsilon_k
\left[1+\frac{\varepsilon_k(\varepsilon-V_g)}
{\varepsilon_k^2-(\varepsilon-V_g)^2 +\Delta^2
}+\frac{(\varepsilon-V_g)^2 -\Delta^2}
{\varepsilon_k^2-(\varepsilon-V_g)^2 +\Delta^2 }\right]\\
 &=&-\frac{1}{2\pi\rho
v_F^2}\left\{\frac{(\varepsilon-V_g)}{2}\ln\left|\frac{(D-\omega_0/2)^2+\Delta^2-(\varepsilon-V_g)^2
}{\omega_0^2/4+\Delta^2-(\varepsilon-V_g)^2 }\right| \right.\\
&&-\left.\sqrt{\Delta^2-(\varepsilon-V_g)^2}
\left[\arctan\frac{D-\omega_0/2}{\sqrt{\Delta^2-(\varepsilon-V_g)^2}}
-\arctan\frac{-\omega_0/2}{\sqrt{\Delta^2-(\varepsilon-V_g)^2}}\right]+D\right\}.\\
\end{array}
\end{equation}
From the above calculations, one readily finds that when the
 gate voltage is taken into account, the Green's functions for the irradiated graphene are
 obtained only by changing $\varepsilon$ to $\varepsilon-V_g$.
\begin{equation}
\begin{array}{cll}
&&\hspace*{-2cm}-\frac{1}{2\pi\rho
v_F^2}\int_{-\frac{\omega_0}{2}}^{D-\omega_0/2} d\varepsilon_k
\frac{\frac{\omega_0}{2}(\varepsilon+\varepsilon_k)}
{\varepsilon_k^2-\varepsilon^2 +\Delta^2 }\\
&\hspace*{-2cm}=&\hspace*{-2cm}-\frac{1}{2\pi\rho
v_F^2}\int_{-\frac{\omega_0}{2}}^{D-\omega_0/2} d\varepsilon_k
\left[\frac{\frac{\omega_0}{2}\varepsilon_k}
{\varepsilon_k^2-\varepsilon^2 +\Delta^2
}+\frac{\frac{\omega_0}{2}\varepsilon}
{\varepsilon_k^2-\varepsilon^2 +\Delta^2 }\right]\\
 &\hspace*{-2cm}=&\hspace*{-2cm}-\frac{1}{2\pi\rho
v_F^2}\left\{\frac{\omega_0}{4}\ln\left|\frac{(D-\omega_0/2)^2+\Delta^2-\varepsilon^2
}{\omega_0^2/4+\Delta^2-\varepsilon^2 }\right|
+\frac{\frac{\omega_0}{2}\varepsilon}{\sqrt{\Delta^2-\varepsilon^2}}
\left[\arctan\frac{D-\omega_0/2}{\sqrt{\Delta^2-\varepsilon^2}}
-\arctan\frac{-\omega_0/2}{\sqrt{\Delta^2-\varepsilon^2}}\right]\right\}.\\
\end{array}
\end{equation}
Thus
\begin{equation}
\begin{array}{cll}
g_{++}^{\tau\tau,r}
 &=&-\frac{1}{2\pi\rho
v_F^2}\left\{(\frac{\varepsilon}{2}+\frac{\omega_0}{4})\ln\left|\frac{(D-\omega_0/2)^2+\Delta^2-\varepsilon^2
}{\omega_0^2/4+\Delta^2-\varepsilon^2 }\right| \right.\\
&&-\left.\frac{1}{\sqrt{\Delta^2-\varepsilon^2}}(\Delta^2-\varepsilon^2-\frac{\omega_0}{2}\varepsilon)
[\arctan\frac{D-\omega_0/2}{\sqrt{\Delta^2-\varepsilon^2}}
-\arctan\frac{-\omega_0/2}{\sqrt{\Delta^2-\varepsilon^2}}]+D\right\},\\
\end{array}
\end{equation}
\begin{equation}
\begin{array}{cll}
&& \hspace*{-2cm}g_{--}^{\tau\tau,r} =-\frac{1}{2\pi\rho
v_F^2}\int_{-\frac{\omega_0}{2}}^{D-\omega_0/2} d\varepsilon_k
\frac{(\varepsilon_k+\frac{\omega_0}{2})(\varepsilon-\varepsilon_k)}
{\varepsilon_k^2-\varepsilon^2 +\Delta^2 }=-\frac{1}{2\pi\rho
v_F^2}\int_{-\frac{\omega_0}{2}}^{D-\omega_0/2} d\varepsilon_k
\left[\frac{\varepsilon_k(\varepsilon-\varepsilon_k)}
{\varepsilon_k^2-\varepsilon^2 +\Delta^2
}+\frac{\frac{\omega_0}{2}(\varepsilon-\varepsilon_k)}
{\varepsilon_k^2-\varepsilon^2 +\Delta^2 }\right]\\
&\hspace*{-2cm}=&\hspace*{-2cm}-\frac{1}{2\pi\rho
v_F^2}\int_{-\frac{\omega_0}{2}}^{D-\omega_0/2} d\varepsilon_k
[\frac{\varepsilon_k\varepsilon}
{\varepsilon_k^2+\Delta^2-\varepsilon^2
}+\frac{\Delta^2-\varepsilon^2 }
{\varepsilon_k^2+\Delta^2-\varepsilon^2
}-1+\frac{\frac{\omega_0}{2}(\varepsilon-\varepsilon_k)}
{\varepsilon_k^2-\varepsilon^2 +\Delta^2 }]\\
&\hspace*{-2cm}=&\hspace*{-2cm}-\frac{1}{2\pi\rho
v_F^2}\int_{-\frac{\omega_0}{2}}^{D-\omega_0/2} d\varepsilon_k
[\frac{\varepsilon_k(\varepsilon-\omega_0/2)}
{\varepsilon_k^2+\Delta^2-\varepsilon^2
}+\frac{\Delta^2-\varepsilon^2+\varepsilon\omega_0/2 }
{\varepsilon_k^2+\Delta^2-\varepsilon^2
}-1]\\
&\hspace*{-2cm}=&\hspace*{-2cm}-\frac{1}{2\pi\rho
v_F^2}\left\{(\frac{\varepsilon}{2}-\frac{\omega_0}{4})\ln\left|\frac{(D-\omega_0/2)^2+\Delta^2-\varepsilon^2
}{\omega_0^2/4+\Delta^2-\varepsilon^2 }\right|
+\frac{\Delta^2-\varepsilon^2+\varepsilon\omega_0/2}{\sqrt{\Delta^2-\varepsilon^2}}
\left[\arctan\frac{D-\omega_0/2}{\sqrt{\Delta^2-\varepsilon^2}}
-\arctan\frac{-\omega_0/2}{\sqrt{\Delta^2-\varepsilon^2}}\right] \right.\\
&&- \left.D\right\},
\label{g22}
\end{array}
\end{equation}
\begin{equation}
\begin{array}{cll}
\hspace*{-2cm}g_{-+}^{\tau\tau,r}=g_{+-}^{\tau\tau,r} =-\frac{1}{2\pi\rho
v_F^2}\int_{-\frac{\omega_0}{2}}^{D-\omega_0/2} d\varepsilon_k
\frac{(\varepsilon_k+\omega_0/2)\Delta}
{\varepsilon_k^2-\varepsilon^2 +\Delta^2 }=-\frac{1}{2\pi\rho
v_F^2}\int_{-\frac{\omega_0}{2}}^{D-\omega_0/2} d\varepsilon_k
\left[\frac{\varepsilon_k \Delta} {\varepsilon_k^2-\varepsilon^2
+\Delta^2 }+\frac{\Delta\omega_0/2}
{\varepsilon_k^2-\varepsilon^2 +\Delta^2 }\right]\\
\hspace*{-2cm}=-\frac{1}{2\pi\rho
v_F^2}\left\{\frac{\Delta}{2}\ln\left|\frac{(D-\omega_0/2)^2+\Delta^2-\varepsilon^2
}{\omega_0^2/4+\Delta^2-\varepsilon^2
}\right|+\frac{\Delta\omega_0/2}{\sqrt{\Delta^2-\varepsilon^2}}[\arctan\frac{D-\omega_0/2}{\sqrt{\Delta^2-\varepsilon^2}}
-\arctan\frac{-\omega_0/2}{\sqrt{\Delta^2-\varepsilon^2}}]\right\}.
\end{array}
\end{equation}

For $|\varepsilon|>\Delta$, we have
\begin{equation}
\begin{array}{cll}
g_{++}^{\tau\tau,r}&=& \frac{1}{N}\sum\limits_{\mathbf{k}}
\frac{\varepsilon-\widetilde{\epsilon}_{-\mathbf{k}}}
{(\varepsilon-\widetilde{\epsilon}_{+\mathbf{k}})
(\varepsilon-\widetilde{\epsilon}_{-\mathbf{k}})-\Delta^2}\\

&=&\frac{S}{(2\pi)^2N}\int
kdkd\phi\frac{\varepsilon+(v_Fk-\frac{\omega_0}{2})}
{\varepsilon^2-(v_Fk-\frac{\omega_0}{2})^2
-\Delta^2}\\

&=&\frac{1}{(2\pi)^2\rho}\frac{2\pi}{v_F^2}\int_{-\frac{\omega_0}{2}}^{D-\omega_0/2}
(\varepsilon_k+\frac{\omega_0}{2})d\varepsilon_k
\frac{\varepsilon+\varepsilon_k}
{\varepsilon^2-\varepsilon_k^2 -\Delta^2+i {\rm sgn}(\varepsilon)\eta}\\
&=&\frac{1}{(2\pi)^2\rho}\frac{2\pi}{v_F^2}\int_{-\frac{\omega_0}{2}}^{D-\omega_0/2}
d\varepsilon_k
\left[\frac{\varepsilon_k(\varepsilon+\varepsilon_k)}
{\varepsilon^2-\varepsilon_k^2
-\Delta^2+i {\rm sgn}(\varepsilon)\eta}

+\frac{\frac{\omega_0}{2}(\varepsilon+\varepsilon_k)}
{\varepsilon^2-\varepsilon_k^2 -\Delta^2+i {\rm sgn}(\varepsilon)\eta}\right]\\

&=&-\frac{1}{(2\pi)^2\rho}\frac{2\pi}{v_F^2}\int_{-\frac{\omega_0}{2}}^{D-\omega_0/2}
d\varepsilon_k
\left[\frac{\varepsilon_k(\varepsilon+\varepsilon_k)}
{\varepsilon_k^2-\varepsilon^2
+\Delta^2-i {\rm sgn}(\varepsilon)\eta}+\frac{\frac{\omega_0}{2}(\varepsilon+\varepsilon_k)}
{\varepsilon_k^2-\varepsilon^2 +\Delta^2-i {\rm sgn}(\varepsilon)\eta}\right].\\

\end{array}
\end{equation}

\begin{equation}
\begin{array}{cll}
g_{--}^{\tau\tau,r}&=& \frac{1}{N}\sum\limits_{\mathbf{k}}
\frac{\varepsilon-\widetilde{\epsilon}_{+\mathbf{k}}}
{(\varepsilon-\widetilde{\epsilon}_{+\mathbf{k}})
(\varepsilon-\widetilde{\epsilon}_{-\mathbf{k}})-\Delta^2}\\

&=&\frac{S}{(2\pi)^2N}\int
kdkd\phi\frac{\varepsilon-(v_Fk-\frac{\omega_0}{2})}
{\varepsilon^2-(v_Fk-\frac{\omega_0}{2})^2
-\Delta^2}\\

&=&\frac{1}{(2\pi)^2\rho}\frac{2\pi}{v_F^2}\int_{-\frac{\omega_0}{2}}^{D-\omega_0/2}
(\varepsilon_k+\frac{\omega_0}{2})d\varepsilon_k
\frac{\varepsilon-\varepsilon_k}
{\varepsilon^2-\varepsilon_k^2 -\Delta^2+i {\rm sgn}(\varepsilon)\eta}\\
&=&\frac{1}{(2\pi)^2\rho}\frac{2\pi}{v_F^2}\int_{-\frac{\omega_0}{2}}^{D-\omega_0/2}
d\varepsilon_k
\left[\frac{\varepsilon_k(\varepsilon-\varepsilon_k)}
{\varepsilon^2-\varepsilon_k^2
-\Delta^2+i {\rm sgn}(\varepsilon)\eta}

+\frac{\frac{\omega_0}{2}(\varepsilon-\varepsilon_k)}
{\varepsilon^2-\varepsilon_k^2 -\Delta^2+i{\rm sgn}(\varepsilon)\eta}\right]\\

&=&-\frac{1}{(2\pi)^2\rho}\frac{2\pi}{v_F^2}\int_{-\frac{\omega_0}{2}}^{D-\omega_0/2}
d\varepsilon_k
\left[\frac{\varepsilon_k(\varepsilon-\varepsilon_k)}
{\varepsilon_k^2-\varepsilon^2
+\Delta^2-i {\rm sgn}(\varepsilon)\eta}+\frac{\frac{\omega_0}{2}(\varepsilon-\varepsilon_k)}
{\varepsilon_k^2-\varepsilon^2 +\Delta^2-i {\rm sgn}(\varepsilon)\eta}\right].\\

\end{array}
\end{equation}

\begin{equation}
\begin{array}{cll}
g_{+-}^{\tau\tau,r}&=&g_{-+}^{\tau\tau,r}=
\frac{1}{N}\sum\limits_{\mathbf{k}} \frac{\Delta}
{(\varepsilon-\widetilde{\epsilon}_{+\mathbf{k}})
(\varepsilon-\widetilde{\epsilon}_{-\mathbf{k}})-\Delta^2}\\

&=&\frac{S}{(2\pi)^2N}\int kdkd\phi\frac{\Delta}
{\varepsilon^2-(v_Fk-\frac{\omega_0}{2})^2
-\Delta^2}\\

&=&\frac{1}{(2\pi)^2\rho}\frac{2\pi}{v_F^2}\int_{-\frac{\omega_0}{2}}^{D-\omega_0/2}
(\varepsilon_k+\frac{\omega_0}{2})d\varepsilon_k \frac{\Delta}
{\varepsilon^2-\varepsilon_k^2 -\Delta^2+i {\rm sgn}(\varepsilon)\eta}\\

&=&\frac{1}{(2\pi)^2\rho}\frac{2\pi}{v_F^2}\int_{-\frac{\omega_0}{2}}^{D-\omega_0/2}
d\varepsilon_k \left[\frac{\varepsilon_k\Delta}
{\varepsilon^2-\varepsilon_k^2
-\Delta^2+i {\rm sgn}(\varepsilon)\eta}

+\frac{\frac{\omega_0}{2}\Delta}
{\varepsilon^2-\varepsilon_k^2 -\Delta^2+i {\rm sgn}(\varepsilon)\eta}\right]\\

&=&-\frac{1}{2\pi\rho v_F^2}
\int_{-\frac{\omega_0}{2}}^{D-\omega_0/2} d\varepsilon_k
\left[\frac{\varepsilon_k\Delta} {\varepsilon_k^2-\varepsilon^2
+\Delta^2-i {\rm sgn}(\varepsilon)\eta}+\frac{\frac{\omega_0}{2}\Delta}
{\varepsilon_k^2-\varepsilon^2 +\Delta^2-i {\rm sgn}(\varepsilon)\eta}\right].\\

\end{array}
\end{equation}

Since
\begin{equation}
\begin{array}{cll}
\hspace*{-1cm}&&\hspace*{-3cm}-\frac{1}{2\pi\rho
v_F^2}\int_{-\frac{\omega_0}{2}}^{D-\omega_0/2} d\varepsilon_k
\frac{\varepsilon_k(\varepsilon+\varepsilon_k)}
{[\varepsilon_k-\sqrt{\varepsilon^2
-\Delta^2}-i {\rm sgn}(\varepsilon)\eta]
[\varepsilon_k+\sqrt{\varepsilon^2 -\Delta^2}+i{\rm sgn}(\varepsilon)\eta ] }\\

&\hspace*{-3cm}=&\hspace*{-3cm}-\frac{1}{4\pi\rho
v_F^2}\int_{-\frac{\omega_0}{2}}^{D-\omega_0/2} d\varepsilon_k
\{\frac{\varepsilon+\varepsilon_k}
{\varepsilon_k+\sqrt{\varepsilon^2
-\Delta^2}+i{\rm sgn}(\varepsilon)\eta  }

+\frac{\varepsilon+\varepsilon_k}
{\varepsilon_k-\sqrt{\varepsilon^2
-\Delta^2}-i{\rm sgn}(\varepsilon)\eta }\}\\

&\hspace*{-3cm}=&\hspace*{-3cm}-\frac{1}{4\pi\rho
v_F^2}\int_{-\frac{\omega_0}{2}}^{D-\omega_0/2} d\varepsilon_k
\{\frac{\varepsilon+\varepsilon_k}
{\varepsilon_k+\sqrt{\varepsilon^2 -\Delta^2} }

+\frac{\varepsilon+\varepsilon_k}
{\varepsilon_k-\sqrt{\varepsilon^2
-\Delta^2} }\}\\

&&\hspace*{-3cm}-\frac{1}{4\pi\rho v_F^2}\{-i\pi{\rm sgn}(\varepsilon)
\theta(\Delta<|\varepsilon|<\sqrt{\frac{\omega_0^2}{4}+\Delta^2})(\varepsilon-\sqrt{\varepsilon^2
-\Delta^2})\\

&&\hspace*{-3cm}+i\pi{\rm sgn}(\varepsilon)
\theta(\Delta<|\varepsilon|<\sqrt{(D-\frac{\omega_0}{2})^2+\Delta^2})(\varepsilon+\sqrt{\varepsilon^2
-\Delta^2})\}\\

&\hspace*{-3cm}=&\hspace*{-3cm}-\frac{1}{4\pi\rho
v_F^2}\int_{-\frac{\omega_0}{2}}^{D-\omega_0/2} d\varepsilon_k
\{\frac{\varepsilon+\varepsilon_k}
{\varepsilon_k+\sqrt{\varepsilon^2 -\Delta^2}   }

+\frac{\varepsilon+\varepsilon_k}
{\varepsilon_k-\sqrt{\varepsilon^2
-\Delta^2}  }\}\\

&&\hspace*{-3cm}+\frac{1}{4\pi\rho v_F^2}i\pi{\rm sgn}(\varepsilon) \{
\theta(\Delta<|\varepsilon|<\sqrt{\frac{\omega_0^2}{4}+\Delta^2})
(\varepsilon-\sqrt{\varepsilon^2 -\Delta^2})\\
&& -
\theta(\Delta<|\varepsilon|<\sqrt{(D-\frac{\omega_0}{2})^2+\Delta^2})(\varepsilon+\sqrt{\varepsilon^2
-\Delta^2})\}\\
\end{array}
\end{equation}

\begin{equation}
\begin{array}{cll}
&&\hspace*{-3cm}-\frac{1}{2\pi\rho
v_F^2}\int_{-\frac{\omega_0}{2}}^{D-\omega_0/2} d\varepsilon_k
\frac{\frac{\omega_0}{2}(\varepsilon+\varepsilon_k)}
{[\varepsilon_k-\sqrt{\varepsilon^2
-\Delta^2}-i{\rm sgn}(\varepsilon)\eta]
[\varepsilon_k+\sqrt{\varepsilon^2 -\Delta^2}+i{\rm sgn}(\varepsilon)\eta ] }\\

&\hspace*{-3cm}=&\hspace*{-3cm}-\frac{1}{4\pi\rho
v_F^2}\int_{-\frac{\omega_0}{2}}^{D-\omega_0/2} d\varepsilon_k
\frac{-\frac{\omega_0}{2}}{\sqrt{\varepsilon^2
-\Delta^2}}\{\frac{\varepsilon+\varepsilon_k}
{\varepsilon_k+\sqrt{\varepsilon^2
-\Delta^2}+i{\rm sgn}(\varepsilon)\eta  }

-\frac{\varepsilon+\varepsilon_k}
{\varepsilon_k-\sqrt{\varepsilon^2
-\Delta^2}-i{\rm sgn}(\varepsilon)\eta }\}\\

&\hspace*{-3cm}=&\hspace*{-3cm}-\frac{1}{4\pi\rho
v_F^2}\int_{-\frac{\omega_0}{2}}^{D-\omega_0/2} d\varepsilon_k
\frac{-\frac{\omega_0}{2}}{\sqrt{\varepsilon^2
-\Delta^2}}\{\frac{\varepsilon+\varepsilon_k}
{\varepsilon_k+\sqrt{\varepsilon^2 -\Delta^2} }

-\frac{\varepsilon+\varepsilon_k}
{\varepsilon_k-\sqrt{\varepsilon^2
-\Delta^2} }\}\\

&&\hspace*{-3cm}-\frac{1}{4\pi\rho
v_F^2}\frac{-\frac{\omega_0}{2}}{\sqrt{\varepsilon^2
-\Delta^2}}\{-i\pi{\rm sgn}(\varepsilon)
\theta(\Delta<|\varepsilon|<\sqrt{\frac{\omega_0^2}{4}+\Delta^2})(\varepsilon-\sqrt{\varepsilon^2
-\Delta^2})\\

&&\hspace*{-3cm}-i\pi{\rm sgn}(\varepsilon)
\theta(\Delta<|\varepsilon|<\sqrt{(D-\frac{\omega_0}{2})^2+\Delta^2})(\varepsilon+\sqrt{\varepsilon^2
-\Delta^2})\}\\

&\hspace*{-3cm}=&\hspace*{-3cm}\frac{1}{4\pi\rho
v_F^2}\frac{\frac{\omega_0}{2}}{\sqrt{\varepsilon^2
-\Delta^2}}\int_{-\frac{\omega_0}{2}}^{D-\omega_0/2}
d\varepsilon_k \{\frac{\varepsilon+\varepsilon_k}
{\varepsilon_k+\sqrt{\varepsilon^2 -\Delta^2}   }

-\frac{\varepsilon+\varepsilon_k}
{\varepsilon_k-\sqrt{\varepsilon^2
-\Delta^2}  }\}\\
\end{array}
\end{equation}
we obtain
\begin{equation}
\begin{array}{cll}
g_{++}^{\tau\tau,r}

&=&-\frac{1}{4\pi\rho v_F^2}\left[

(\varepsilon+\frac{\frac{\omega_0}{2}}{\sqrt{\varepsilon^2
-\Delta^2}}\sqrt{\varepsilon^2
-\Delta^2})\ln\left|\frac{(D-\omega_0/2)^2-\varepsilon^2
+\Delta^2}{\varepsilon^2 -\Delta^2-\omega_0^2/4}\right|\right.\\

&&\left.-(\varepsilon\frac{\frac{\omega_0}{2}}{\sqrt{\varepsilon^2
-\Delta^2}}+\sqrt{\varepsilon^2 -\Delta^2})
\ln\left|\frac{(D-\omega_0/2+\sqrt{\varepsilon^2
-\Delta^2})(\omega_0/2+\sqrt{\varepsilon^2
-\Delta^2})}{(D-\omega_0/2-\sqrt{\varepsilon^2
-\Delta^2})(\omega_0/2-\sqrt{\varepsilon^2 -\Delta^2})}\right|+2D
\right]\\

&&+\frac{1}{4\pi\rho v_F^2}i\pi{\rm sgn}(\varepsilon) \{
\theta(\Delta<|\varepsilon|<\sqrt{\frac{\omega_0^2}{4}+\Delta^2})
(\varepsilon-\sqrt{\varepsilon^2
-\Delta^2})(1-\frac{\frac{\omega_0}{2}}{\sqrt{\varepsilon^2 -\Delta^2}})\\
&& -
\theta(\Delta<|\varepsilon|<\sqrt{(D-\frac{\omega_0}{2})^2+\Delta^2})(\varepsilon+\sqrt{\varepsilon^2
-\Delta^2})(1+\frac{\frac{\omega_0}{2}}{\sqrt{\varepsilon^2 -\Delta^2}})\}.\\

\end{array}
\end{equation}
In order to calculate $g_{--}^{\tau\tau,r}$, we  solve at first  for
\begin{equation}
\begin{array}{cll}
&&\hspace*{-3cm}-\frac{1}{2\pi\rho
v_F^2}\int_{-\frac{\omega_0}{2}}^{D-\omega_0/2} d\varepsilon_k
\frac{\varepsilon_k(\varepsilon-\varepsilon_k)}
{[\varepsilon_k-\sqrt{\varepsilon^2
-\Delta^2}-i{\rm sgn}(\varepsilon)\eta]
[\varepsilon_k+\sqrt{\varepsilon^2 -\Delta^2}+i{\rm sgn}(\varepsilon)\eta ] }\\

&\hspace*{-3cm}=&\hspace*{-3cm}-\frac{1}{4\pi\rho
v_F^2}\int_{-\frac{\omega_0}{2}}^{D-\omega_0/2} d\varepsilon_k
\{\frac{\varepsilon-\varepsilon_k}
{\varepsilon_k+\sqrt{\varepsilon^2
-\Delta^2}+i{\rm sgn}(\varepsilon)\eta  }

+\frac{\varepsilon-\varepsilon_k}
{\varepsilon_k-\sqrt{\varepsilon^2
-\Delta^2}-i{\rm sgn}(\varepsilon)\eta }\}\\

&\hspace*{-3cm}=&\hspace*{-3cm}-\frac{1}{4\pi\rho
v_F^2}\int_{-\frac{\omega_0}{2}}^{D-\omega_0/2} d\varepsilon_k
\{\frac{\varepsilon-\varepsilon_k}
{\varepsilon_k+\sqrt{\varepsilon^2 -\Delta^2} }

+\frac{\varepsilon-\varepsilon_k}
{\varepsilon_k-\sqrt{\varepsilon^2
-\Delta^2} }\}\\

&&\hspace*{-3cm}-\frac{1}{4\pi\rho v_F^2}\{-i\pi{\rm sgn}(\varepsilon)
\theta(\Delta<|\varepsilon|<\sqrt{\frac{\omega_0^2}{4}+\Delta^2})(\varepsilon+\sqrt{\varepsilon^2
-\Delta^2})\\

&&\hspace*{-3cm}+i\pi{\rm sgn}(\varepsilon)
\theta(\Delta<|\varepsilon|<\sqrt{(D-\frac{\omega_0}{2})^2+\Delta^2})(\varepsilon-\sqrt{\varepsilon^2
-\Delta^2})\}\\

&\hspace*{-3cm}=&\hspace*{-3cm}-\frac{1}{4\pi\rho
v_F^2}\int_{-\frac{\omega_0}{2}}^{D-\omega_0/2} d\varepsilon_k
\{\frac{\varepsilon-\varepsilon_k}
{\varepsilon_k+\sqrt{\varepsilon^2 -\Delta^2}   }

+\frac{\varepsilon-\varepsilon_k}
{\varepsilon_k-\sqrt{\varepsilon^2
-\Delta^2}  }\}\\

&&\hspace*{-3cm}+\frac{1}{4\pi\rho v_F^2}i\pi{\rm sgn}(\varepsilon) \{
\theta(\Delta<|\varepsilon|<\sqrt{\frac{\omega_0^2}{4}+\Delta^2})
(\varepsilon+\sqrt{\varepsilon^2 -\Delta^2})\\
&&\hspace*{-3cm} -
\theta(\Delta<|\varepsilon|<\sqrt{(D-\frac{\omega_0}{2})^2+\Delta^2})(\varepsilon-\sqrt{\varepsilon^2
-\Delta^2})\}\\
\end{array}
\end{equation}

\begin{equation}
\begin{array}{cll}
&&\hspace*{-3cm}-\frac{1}{2\pi\rho
v_F^2}\int_{-\frac{\omega_0}{2}}^{D-\omega_0/2} d\varepsilon_k
\frac{\frac{\omega_0}{2}(\varepsilon-\varepsilon_k)}
{[\varepsilon_k-\sqrt{\varepsilon^2
-\Delta^2}-i{\rm sgn}(\varepsilon)\eta]
[\varepsilon_k+\sqrt{\varepsilon^2 -\Delta^2}+i{\rm sgn}(\varepsilon)\eta ] }\\

&\hspace*{-3cm}=&\hspace*{-3cm}-\frac{1}{4\pi\rho
v_F^2}\int_{-\frac{\omega_0}{2}}^{D-\omega_0/2} d\varepsilon_k
\frac{-\frac{\omega_0}{2}}{\sqrt{\varepsilon^2
-\Delta^2}}\{\frac{\varepsilon-\varepsilon_k}
{\varepsilon_k+\sqrt{\varepsilon^2
-\Delta^2}+i{\rm sgn}(\varepsilon)\eta  }

-\frac{\varepsilon-\varepsilon_k}
{\varepsilon_k-\sqrt{\varepsilon^2
-\Delta^2}-i{\rm sgn}(\varepsilon)\eta }\}\\

&\hspace*{-3cm}=&\hspace*{-3cm}-\frac{1}{4\pi\rho
v_F^2}\int_{-\frac{\omega_0}{2}}^{D-\omega_0/2} d\varepsilon_k
\frac{-\frac{\omega_0}{2}}{\sqrt{\varepsilon^2
-\Delta^2}}\{\frac{\varepsilon-\varepsilon_k}
{\varepsilon_k+\sqrt{\varepsilon^2 -\Delta^2} }

-\frac{\varepsilon-\varepsilon_k}
{\varepsilon_k-\sqrt{\varepsilon^2
-\Delta^2} }\}\\

&&\hspace*{-3cm}-\frac{1}{4\pi\rho
v_F^2}\frac{-\frac{\omega_0}{2}}{\sqrt{\varepsilon^2
-\Delta^2}}\{-i\pi{\rm sgn}(\varepsilon)
\theta(\Delta<|\varepsilon|<\sqrt{\frac{\omega_0^2}{4}+\Delta^2})(\varepsilon+\sqrt{\varepsilon^2
-\Delta^2})\\

&&\hspace*{-3cm}-i\pi{\rm sgn}(\varepsilon)
\theta(\Delta<|\varepsilon|<\sqrt{(D-\frac{\omega_0}{2})^2+\Delta^2})(\varepsilon-\sqrt{\varepsilon^2
-\Delta^2})\}\\

&\hspace*{-3cm}=&\hspace*{-3cm}\frac{1}{4\pi\rho
v_F^2}\frac{\frac{\omega_0}{2}}{\sqrt{\varepsilon^2
-\Delta^2}}\int_{-\frac{\omega_0}{2}}^{D-\omega_0/2}
d\varepsilon_k \{\frac{\varepsilon-\varepsilon_k}
{\varepsilon_k+\sqrt{\varepsilon^2 -\Delta^2}   }

-\frac{\varepsilon-\varepsilon_k}
{\varepsilon_k-\sqrt{\varepsilon^2
-\Delta^2}  }\}\\
\end{array}
\end{equation}

This means that

\begin{equation}
\begin{array}{cll}
\hspace*{-3cm}g_{--}^{\tau\tau,r} &=&-\frac{1}{4\pi\rho v_F^2}\left[

(\varepsilon-\frac{\frac{\omega_0}{2}}{\sqrt{\varepsilon^2
-\Delta^2}}\sqrt{\varepsilon^2
-\Delta^2})\ln\left|\frac{(D-\omega_0/2)^2-\varepsilon^2
+\Delta^2}{\varepsilon^2 -\Delta^2-\omega_0^2/4}\right|\right.\\

&&\left.+(\sqrt{\varepsilon^2
-\Delta^2}-\varepsilon\frac{\frac{\omega_0}{2}}{\sqrt{\varepsilon^2
-\Delta^2}}) \ln\left|\frac{(D-\omega_0/2+\sqrt{\varepsilon^2
-\Delta^2})(\omega_0/2+\sqrt{\varepsilon^2
-\Delta^2})}{(D-\omega_0/2-\sqrt{\varepsilon^2
-\Delta^2})(\omega_0/2-\sqrt{\varepsilon^2 -\Delta^2})}\right|-2D
\right]\\

&&+\frac{1}{4\pi\rho v_F^2}i\pi{\rm sgn}(\varepsilon) \{
\theta(\Delta<|\varepsilon|<\sqrt{\frac{\omega_0^2}{4}+\Delta^2})
(\varepsilon+\sqrt{\varepsilon^2
-\Delta^2})(1-\frac{\frac{\omega_0}{2}}{\sqrt{\varepsilon^2 -\Delta^2}})\\
&& -
\theta(\Delta<|\varepsilon|<\sqrt{(D-\frac{\omega_0}{2})^2+\Delta^2})(\varepsilon-\sqrt{\varepsilon^2
-\Delta^2})(1+\frac{\frac{\omega_0}{2}}{\sqrt{\varepsilon^2 -\Delta^2}})\}.\\

\end{array}
\end{equation}
Similar to the calculations of the Green's functions
$g_{++}^{\tau\tau,r}$ and $g_{--}^{\tau\tau,r}$, we  consider at first
the following terms in $g_{+-}^{\tau\tau,r}$,
\begin{equation}
\begin{array}{cll}
&&\hspace*{-3cm}-\frac{1}{2\pi\rho
v_F^2}\int_{-\frac{\omega_0}{2}}^{D-\omega_0/2} d\varepsilon_k
\frac{\varepsilon_k\Delta} {[\varepsilon_k-\sqrt{\varepsilon^2
-\Delta^2}-i{\rm sgn}(\varepsilon)\eta]
[\varepsilon_k+\sqrt{\varepsilon^2 -\Delta^2}+i{\rm sgn}(\varepsilon)\eta ] }\\

&\hspace*{-3cm}=&\hspace*{-3cm}-\frac{1}{4\pi\rho
v_F^2}\int_{-\frac{\omega_0}{2}}^{D-\omega_0/2} d\varepsilon_k
\{\frac{\Delta} {\varepsilon_k+\sqrt{\varepsilon^2
-\Delta^2}+i{\rm sgn}(\varepsilon)\eta  }

+\frac{\Delta} {\varepsilon_k-\sqrt{\varepsilon^2
-\Delta^2}-i{\rm sgn}(\varepsilon)\eta }\}\\

&\hspace*{-3cm}=&\hspace*{-3cm}-\frac{1}{4\pi\rho
v_F^2}\int_{-\frac{\omega_0}{2}}^{D-\omega_0/2} d\varepsilon_k
\{\frac{\Delta} {\varepsilon_k+\sqrt{\varepsilon^2 -\Delta^2} }

+\frac{\Delta} {\varepsilon_k-\sqrt{\varepsilon^2
-\Delta^2} }\}\\

&&\hspace*{-3cm}-\frac{1}{4\pi\rho v_F^2}\{-i\pi{\rm sgn}(\varepsilon)
\theta(\Delta<|\varepsilon|<\sqrt{\frac{\omega_0^2}{4}+\Delta^2})\Delta\\

&&\hspace*{-3cm}+i\pi{\rm sgn}(\varepsilon)
\theta(\Delta<|\varepsilon|<\sqrt{(D-\frac{\omega_0}{2})^2+\Delta^2})\Delta\}\\

&\hspace*{-3cm}=&\hspace*{-3cm}-\frac{1}{4\pi\rho
v_F^2}\int_{-\frac{\omega_0}{2}}^{D-\omega_0/2} d\varepsilon_k
\{\frac{\Delta} {\varepsilon_k+\sqrt{\varepsilon^2 -\Delta^2}   }

+\frac{\Delta} {\varepsilon_k-\sqrt{\varepsilon^2
-\Delta^2}  }\}\\

&&\hspace*{-3cm}+\frac{\Delta}{4\pi\rho v_F^2}i\pi{\rm sgn}(\varepsilon) \{
\theta(\Delta<|\varepsilon|<\sqrt{\frac{\omega_0^2}{4}+\Delta^2})
  -
\theta(\Delta<|\varepsilon|<\sqrt{(D-\frac{\omega_0}{2})^2+\Delta^2})\}\\

&\hspace*{-3cm}=&\hspace*{-3cm}-\frac{\Delta}{4\pi\rho v_F^2}
\ln\left|\frac{(D-\omega_0/2)^2-\varepsilon^2
+\Delta^2}{\varepsilon^2 -\Delta^2-\omega_0^2/4}\right|

\\

&&\hspace*{-3cm}+\frac{\Delta}{4\pi\rho v_F^2}i\pi{\rm sgn}(\varepsilon) \{
\theta(\Delta<|\varepsilon|<\sqrt{\frac{\omega_0^2}{4}+\Delta^2})
  -
\theta(\Delta<|\varepsilon|<\sqrt{(D-\frac{\omega_0}{2})^2+\Delta^2})\}\\
\end{array}
\end{equation}

\begin{equation}
\begin{array}{cll}
&&\hspace*{-3cm}-\frac{1}{2\pi\rho
v_F^2}\int_{-\frac{\omega_0}{2}}^{D-\omega_0/2} d\varepsilon_k
\frac{\frac{\omega_0}{2}\Delta}
{[\varepsilon_k-\sqrt{\varepsilon^2
-\Delta^2}-i{\rm sgn}(\varepsilon)\eta]
[\varepsilon_k+\sqrt{\varepsilon^2 -\Delta^2}+i{\rm sgn}(\varepsilon)\eta ] }\\

&\hspace*{-3cm}=&\hspace*{-3cm}-\frac{1}{4\pi\rho
v_F^2}\int_{-\frac{\omega_0}{2}}^{D-\omega_0/2} d\varepsilon_k
\frac{-\frac{\omega_0}{2}}{\sqrt{\varepsilon^2
-\Delta^2}}\{\frac{\Delta} {\varepsilon_k+\sqrt{\varepsilon^2
-\Delta^2}+i{\rm sgn}(\varepsilon)\eta  }

-\frac{\Delta} {\varepsilon_k-\sqrt{\varepsilon^2
-\Delta^2}-i{\rm sgn}(\varepsilon)\eta }\}\\

&\hspace*{-3cm}=&\hspace*{-3cm}-\frac{1}{4\pi\rho
v_F^2}\int_{-\frac{\omega_0}{2}}^{D-\omega_0/2} d\varepsilon_k
\frac{-\frac{\omega_0}{2}}{\sqrt{\varepsilon^2
-\Delta^2}}\{\frac{\Delta} {\varepsilon_k+\sqrt{\varepsilon^2
-\Delta^2} }

-\frac{\Delta} {\varepsilon_k-\sqrt{\varepsilon^2
-\Delta^2} }\}\\

&&\hspace*{-3cm}-\frac{1}{4\pi\rho
v_F^2}\frac{-\frac{\omega_0}{2}}{\sqrt{\varepsilon^2
-\Delta^2}}\{-i\pi{\rm sgn}(\varepsilon)
\theta(\Delta<|\varepsilon|<\sqrt{\frac{\omega_0^2}{4}+\Delta^2})\Delta\\

&&\hspace*{-3cm}-i\pi{\rm sgn}(\varepsilon)
\theta(\Delta<|\varepsilon|<\sqrt{(D-\frac{\omega_0}{2})^2+\Delta^2})\Delta\}\\

&\hspace*{-3cm}=&\hspace*{-3cm}\frac{1}{4\pi\rho
v_F^2}\frac{\frac{\omega_0}{2}}{\sqrt{\varepsilon^2
-\Delta^2}}\int_{-\frac{\omega_0}{2}}^{D-\omega_0/2}
d\varepsilon_k \{\frac{\Delta} {\varepsilon_k+\sqrt{\varepsilon^2
-\Delta^2}   }

-\frac{\Delta} {\varepsilon_k-\sqrt{\varepsilon^2
-\Delta^2}  }\}\\

&&\hspace*{-3cm}+\frac{1}{4\pi\rho v_F^2}i\pi{\rm sgn}(\varepsilon)
\frac{-\frac{\omega_0}{2}\Delta}{\sqrt{\varepsilon^2 -\Delta^2}}\{
\theta(\Delta<|\varepsilon|<\sqrt{\frac{\omega_0^2}{4}+\Delta^2})
  +
\theta(\Delta<|\varepsilon|<\sqrt{(D-\frac{\omega_0}{2})^2+\Delta^2})\}\\
\end{array}
\end{equation}

\begin{equation}
\begin{array}{cll}
\hspace*{-3cm} g_{+-}^{\tau\tau,r}&=&g_{-+}^{\tau\tau,r}\\

&\hspace*{-3cm}=&\hspace*{-3cm}-\frac{\Delta}{4\pi\rho v_F^2}
\ln\left|\frac{(D-\omega_0/2)^2-\varepsilon^2
+\Delta^2}{\varepsilon^2 -\Delta^2-\omega_0^2/4}\right|

\\

&&\hspace*{-3cm}+\frac{\Delta}{4\pi\rho v_F^2}i\pi{\rm sgn}(\varepsilon) \{
\theta(\Delta<|\varepsilon|<\sqrt{\frac{\omega_0^2}{4}+\Delta^2})
  -
\theta(\Delta<|\varepsilon|<\sqrt{(D-\frac{\omega_0}{2})^2+\Delta^2})\}\\

&&\hspace*{-3cm}+\frac{1}{4\pi\rho
v_F^2}\frac{\frac{\omega_0}{2}\Delta}{\sqrt{\varepsilon^2
-\Delta^2}} \ln\left|\frac{(D-\omega_0/2+\sqrt{\varepsilon^2
-\Delta^2})(\omega_0/2+\sqrt{\varepsilon^2
-\Delta^2})}{(D-\omega_0/2-\sqrt{\varepsilon^2
-\Delta^2})(\omega_0/2-\sqrt{\varepsilon^2 -\Delta^2})}\right|
\\

&&\hspace*{-3cm}-\frac{\Delta}{4\pi\rho v_F^2}i\pi{\rm sgn}(\varepsilon)
\frac{\frac{\omega_0}{2}}{\sqrt{\varepsilon^2 -\Delta^2}}\{
\theta(\Delta<|\varepsilon|<\sqrt{\frac{\omega_0^2}{4}+\Delta^2})
  +
\theta(\Delta<|\varepsilon|<\sqrt{(D-\frac{\omega_0}{2})^2+\Delta^2})\}\\

&\hspace*{-3cm}=&\hspace*{-3cm}-\frac{\Delta}{4\pi\rho v_F^2}
\left[\ln\left|\frac{(D-\omega_0/2)^2-\varepsilon^2
+\Delta^2}{\varepsilon^2 -\Delta^2-\omega_0^2/4}\right|

-\frac{\frac{\omega_0}{2} }{\sqrt{\varepsilon^2 -\Delta^2}}
\ln\left|\frac{(D-\omega_0/2+\sqrt{\varepsilon^2
-\Delta^2})(\omega_0/2+\sqrt{\varepsilon^2
-\Delta^2})}{(D-\omega_0/2-\sqrt{\varepsilon^2
-\Delta^2})(\omega_0/2-\sqrt{\varepsilon^2
-\Delta^2})}\right|\right]

\\

&&\hspace*{-3cm}+\frac{\Delta}{4\pi\rho v_F^2}i\pi{\rm sgn}(\varepsilon) \{
\theta(\Delta<|\varepsilon|<\sqrt{\frac{\omega_0^2}{4}+\Delta^2})(1-\frac{\frac{\omega_0}{2}
}{\sqrt{\varepsilon^2 -\Delta^2}})\\

&&\hspace*{-3cm}  -
\theta(\Delta<|\varepsilon|<\sqrt{(D-\frac{\omega_0}{2})^2+\Delta^2})(1+\frac{\frac{\omega_0}{2} }{\sqrt{\varepsilon^2 -\Delta^2}})\}.\\

\end{array}
\end{equation}

\subsection{Fourier-transformed Green's functions}

Since $H(t+T)=H(t)$ applies, the  retarded Green's function obeys
$\tilde{G}(t,t')=\tilde{G}(t+T,t'+T)$ (S. Kohler, \emph{et al.},Physics Reports \textbf{406}, 379(2005)). As we can write
$\tilde{G}(t,t')=\tilde{G}(t+T,t'+T)=\tilde{G}(t+T, t+T-(t-t'))$ we
define  $\tau=t$ and $\tau'=t-t'$ and express the Green's functions  as
$\tilde{G}(t+T, t+T-(t-t'))\equiv G(\tau+T,\tau')$,
which satisfies
$G(\tau+T,\tau')=G(\tau,\tau')$.
Thus, one can introduce the Fourier transformation

\begin{equation}
\begin{array}{cll}
G(\tau,\tau')&=&\sum\limits_n e^{i\frac{n\omega_0 \tau}{2}}
\int \frac{d\varepsilon}{2\pi}
e^{-i\varepsilon\tau'}G_n(\varepsilon)\\
\end{array}\label{getranp0}
\end{equation}
which is the same as the generalized Fourier form
\begin{equation}
\begin{array}{cll}
\hspace*{-3cm}G(t,t')=\frac{1}{2\pi}\sum\limits_n\int d\varepsilon
e^{-i\varepsilon
t}e^{i(\varepsilon+n\omega_0/2)t'}G(\varepsilon,\varepsilon+n\omega_0/2)
=\sum\limits_n e^{i\frac{n\omega_0 t}{2}}
\int \frac{d\varepsilon}{2\pi}
e^{-i(\varepsilon+n\omega_0/2)(t-t')}G(\varepsilon,\varepsilon+n\omega_0/2)\\


\hspace*{-11cm}=\sum\limits_n e^{i\frac{n\omega_0 t}{2}}
\int \frac{d\varepsilon}{2\pi}
e^{-i\varepsilon(t-t')}G(\varepsilon- n\omega_0/2,\varepsilon)\\
\end{array}\label{getran}
\end{equation}
For $\tau=t$ and $\tau'=t-t'$ same expression as Eq.(\ref{getranp0}) follows.

Denoting
$G_{mn}(\varepsilon)=G(\varepsilon+m\omega_0/2,\varepsilon
+n\omega_0/2)$ allows to write for the different Fourier components  $G_{mn}(\varepsilon)=G_{m-n,0}(\varepsilon+n\omega_0/2)$, for
%
$G_{mn}(\varepsilon)=G(\varepsilon+m\omega_0/2,\varepsilon+n\omega_0/2)
=G[\varepsilon+n\omega_0/2+(m-n)\omega_0/2,\varepsilon+n\omega_0/2]
=G_{m-n,0}(\varepsilon+n\omega_0/2)
$ applies.
%
For the inverse transformation of Eq.(\ref{getranp0}) one finds
\begin{equation}
\begin{array}{cll}
G_{mn}(\varepsilon)&=&\frac{1}{T}\int dtdt'
e^{i(m\omega_0t/2-n\omega_0t'/2)}e^{i\varepsilon(t-t')}G(t,t').\\
\end{array}\label{getran1}
\end{equation}

For the Fourier transformation of the Green's function  one finds
\begin{equation}
%
\label{smg1}
\end{equation}
For the selfenergy we write
\begin{equation}
%
\end{equation}

The last terms of Eq.(\ref{gremat}) can be cast as

\begin{equation}
\begin{array}{cll}
 &&\hspace*{-3cm}\int dt_1\int dt_2
 g^{\tau\tau,r}(t-t_1)\Sigma^{\tau,r}(t_1,t_2)
G^{\tau\tau',r}(t_2,t')\\

&\hspace*{-3cm}=&\hspace*{-3cm}\frac{1}{(2\pi)^3 }\int dt_1 dt_2\sum\limits_{n_1n_2n_3}\int
d\varepsilon_1 d\varepsilon_2d\varepsilon_3
e^{-i(\varepsilon_1+\frac{m_1\omega_0}{2})t}e^{i(\varepsilon_1+\frac{n_1\omega_0}{2})t_1}
e^{-i(\varepsilon_2+\frac{m_2\omega_0}{2})t_1}e^{i(\varepsilon_2+\frac{n_2\omega_0}{2})t_2}\\
&&\hspace*{-3cm}\times
e^{-i(\varepsilon_3+\frac{m_3\omega_0}{2})t_2}e^{i(\varepsilon_3+\frac{n_3\omega_0}{2})t'}
 g_{m_1n_1}^{\tau\tau,r}(\varepsilon_1)\Sigma_{m_2n_2}^{\tau,r}(\varepsilon_2)
G_{m_3n_3}^{\tau\tau',r}(\varepsilon_3)\\

&\hspace*{-3cm}=&\hspace*{-3cm}\frac{1}{2\pi} \sum\limits_{n_1n_2n_3}\int d\varepsilon_1
d\varepsilon_2d\varepsilon_3
e^{-i(\varepsilon_1+\frac{m_1\omega_0}{2})t}\delta(\varepsilon_1+\frac{n_1\omega_0}{2}-
\varepsilon_2-\frac{m_2\omega_0}{2})\delta(\varepsilon_2+\frac{n_2\omega_0}{2}-
\varepsilon_3-\frac{m_3\omega_0}{2})\\

&&\hspace*{-3cm}\times e^{i(\varepsilon_3+\frac{n_3\omega_0}{2})t'}
 g_{m_1n_1}^{\tau\tau,r}(\varepsilon_1)\Sigma_{m_2n_2}^{\tau,r}(\varepsilon_2)
G_{m_3n_3}^{\tau\tau',r}(\varepsilon_3)\\

&\hspace*{-3cm}=&\hspace*{-3cm}\frac{1}{2\pi} \sum\limits_{n_1n_2n_3}\int
d\varepsilon_2d\varepsilon_3
e^{-i(\varepsilon_2+\frac{m_2\omega_0}{2}-\frac{n_1\omega_0}{2}+\frac{m_1\omega_0}{2})t}
\delta(\varepsilon_2+\frac{n_2\omega_0}{2}-
\varepsilon_3-\frac{m_3\omega_0}{2})\\

&&\hspace*{-3cm}\times e^{i(\varepsilon_3+\frac{n_3\omega_0}{2})t'}
 g_{m_1n_1}^{\tau\tau,r}(\varepsilon_2+\frac{m_2\omega_0}{2}-\frac{n_1\omega_0}{2})
 \Sigma_{m_2n_2}^{\tau,r}(\varepsilon_2)
G_{m_3n_3}^{\tau\tau',r}(\varepsilon_3)\\

&\hspace*{-3cm}=&\hspace*{-3cm}\frac{1}{2\pi} \sum\limits_{n_1n_2n_3} \int d\varepsilon_2
e^{-i(\varepsilon_2+\frac{m_2\omega_0}{2}-\frac{n_1\omega_0}{2}+\frac{m_1\omega_0}{2})t}
e^{i(\varepsilon_2+\frac{n_2\omega_0}{2}-
\frac{m_3\omega_0}{2}+\frac{n_3\omega_0}{2})t'}\\
\end{array}
\end{equation}
Replacing $\varepsilon_2-\frac{n_1\omega_0}{2}$ by $\varepsilon_2$ we arrive at
\begin{equation}
\begin{array}{cll}
 &&\int dt_1\int dt_2
 g^{\tau\tau,r}(t-t_1)\Sigma^{\tau,r}(t_1,t_2)
G^{\tau\tau',r}(t_2,t')\\

&=&\frac{1}{2\pi} \sum\limits_{n_1n_2n_3} \int d\varepsilon_2
e^{-i[\varepsilon_2+(m_2+m_1)\frac{\omega_0}{2}]t}
e^{i[\varepsilon_2+(n_1+n_2-m_3+n_3)\frac{\omega_0}{2}]t'}\\

&&\times
 g_{m_1n_1}^{\tau\tau,r}(\varepsilon_2+m_2\frac{\omega_0}{2})
 \Sigma_{m_2n_2}^{\tau,r}(\varepsilon_2+n_1\frac{\omega_0}{2})
G_{m_3n_3}^{\tau\tau',r}[\varepsilon_2+(n_1+n_2-m_3)\frac{\omega_0}{2}].\\
\end{array}
\end{equation}
Let $m_2=k_1 -m_1$ and $n_3=k_2-n_1-n_2+m_3$, then
\begin{equation}
\begin{array}{cll}
 &&\int dt_1\int dt_2
 g^{\tau\tau,r}(t-t_1)\Sigma^{\tau,r}(t_1,t_2)
G^{\tau\tau',r}(t_2,t')\\
&=&\frac{1}{2\pi} \sum\limits_{n_1n_2k_2} \int d\varepsilon_2
e^{-i(\varepsilon_2+k_1\frac{\omega_0}{2})t}
e^{i(\varepsilon_2+k_2\frac{\omega_0}{2})t'}\\

&&\times
 g_{m_1n_1}^{\tau\tau,r}[\varepsilon_2+(k_1 -m_1)\frac{\omega_0}{2}]
 \Sigma_{k_1 -m_1,n_2}^{\tau,r}(\varepsilon_2+n_1\frac{\omega_0}{2})\nonumber\\ &&
G_{m_3,k_2-n_1-n_2+m_3}^{\tau\tau',r}[\varepsilon_2+(n_1+n_2-m_3)\frac{\omega_0}{2}]\\

&=&\frac{1}{2\pi} \sum\limits_{n_1n_2k_2} \int d\varepsilon_2
e^{-i(\varepsilon_2+k_1\frac{\omega_0}{2})t}
e^{i(\varepsilon_2+k_2\frac{\omega_0}{2})t'}\\

&&\times
 g_{k_1,n_1+k_1 -m_1}^{\tau\tau,r}(\varepsilon_2)
 \Sigma_{n_1+k_1 -m_1,n_1+n_2}^{\tau,r}(\varepsilon_2)
G_{n_1+n_2,k_2}^{\tau\tau',r}(\varepsilon_2)\\

&=&\frac{1}{2\pi} \sum\limits_{n_1n_2k_2}\int d\varepsilon_2
e^{-i(\varepsilon_2+k_1\frac{\omega_0}{2})t}
e^{i(\varepsilon_2+k_2\frac{\omega_0}{2})t'}
 g_{k_1,n_1}^{\tau\tau,r}(\varepsilon_2)
 \Sigma_{n_1,n_2}^{\tau,r}(\varepsilon_2)
G_{n_2,k_2}^{\tau\tau',r}(\varepsilon_2).\\
\end{array}\label{gremat11}
\end{equation}

Substituting Eq.(\ref{gremat11}) in Eq.(\ref{gremat}) we find  the expression
\begin{equation}
\begin{array}{cll}
\mathcal{G}_{mn}^{\tau\tau',r}(\varepsilon)

&=&\delta_{\tau\tau'} g_{mn}^{\tau\tau,r}(\varepsilon) +
 \sum\limits_{n_1n_2}
 g_{m,n_1}^{\tau\tau,r}(\varepsilon)
 \Sigma_{n_1,n_2}^{\tau,r}(\varepsilon)
\mathcal{G}_{n_2,n}^{\tau\tau',r}(\varepsilon)\\
\end{array}\label{grematf}
\end{equation}
The corresponding matrix equations read
\begin{equation}
%
\label{grematf1}
\end{equation}
This means that
\begin{equation}
%
\label{gremtx1}
\end{equation}

Eq.(\ref{gremtx1}) is simply written as
\begin{equation}
\begin{array}{cll}
 \mathcal{G}_{m,n}=g_{mn}+\widetilde{\epsilon}_m\mathcal{G}_{m,n}+V_{m,m-2}\mathcal{G}_{m-2,n}
 +V_{m,m+2}\mathcal{G}_{m+2,n}
\end{array}\label{gremx1}
\end{equation}
with  $g$ being diagonal. Furthermore, we deduce that

%
\begin{equation}
\begin{array}{cll}
 \mathcal{G}_{N,n}=\widetilde{\epsilon}_N\mathcal{G}_{N,n}+V_{N,N-2}\mathcal{G}_{N-2,n},\
 \ N>n
\end{array}\label{ghh6}
\end{equation}
which means that
\begin{equation}
\begin{array}{cll}
 \mathcal{G}_{N,n}=B_N \mathcal{G}_{N-2,n},\ \
 B_N=(I-\widetilde{\epsilon}_N)^{-1}V_{N,N-2}.\
 \ N>n
\end{array}
\end{equation}
Furthermore, it is straightforward to see that
\begin{equation}
\mathcal{G}_{N-2,n}=B_{N-2}G_{N-4,n}^r,\ \
B_{N-2}=[I-\widetilde{\epsilon}_{N-2}-V_{N-2,N}B_N]^{-1}V_{N-2,N-4},\
\ N-1>n
\end{equation}
\begin{equation}
\begin{array}{cll}
 \mathcal{G}_{m,n}=B_{m} \mathcal{G}_{m-2,n},\ \
 B_{m}=(I-\widetilde{\epsilon}_m-V_{m,m+2}B_{m+2})^{-1}V_{m,m-2}.\
 \ m>n
\end{array}\label{mgen}
\end{equation}
Likewise we deduce that
\begin{equation}
\begin{array}{cll}
 \mathcal{G}_{-N,n}=F_{-N}\mathcal{G}_{-N+2,n},\ \
 F_{-N}=(I-\widetilde{\epsilon}_{-N})^{-1}V_{-N,-N+2},
\end{array}
\end{equation}
\begin{equation}
\begin{array}{cll}
 \mathcal{G}_{-N+2,n}=F_{-N+2}\mathcal{G}_{-N+4,n},\ \
 F_{-N+2}\nonumber\\
 =(I-\widetilde{\epsilon}_{-N+2}-V_{-N+2,-N}F_{-N})^{-1}V_{-N+2,-N+4},
\end{array}
\end{equation}
and
\begin{equation}
\begin{array}{cll}
 \mathcal{G}_{m,n}=F_m\mathcal{G}_{m+2,n},\ \
 F_m=(I-\widetilde{\epsilon}_m-V_{m,m-2}F_{m-2})^{-1}V_{m,m+2}.\ \ m<n
\end{array}\label{mlen}
\end{equation}
Using
$\mathcal{G}_{m,n}(\varepsilon)=\mathcal{G}_{m-n,0}(\varepsilon+n\omega_0/2)$ from
 $\mathcal{G}_{m,0}(\varepsilon)$ all the elements of the
 matrix $\mathcal{G}_{mn}(\varepsilon)$ follow.
Denoting
$\mathcal{G}_{m0}(\varepsilon)\equiv \mathcal{G}_m$, i.e.
$ \mathcal{G}_{0}=g_{00}+\widetilde{\epsilon}_0\mathcal{G}_{0}+V_{0,-2}\mathcal{G}_{-2}
 +V_{0,2}\mathcal{G}_{2}
$
and combining Eqs.(\ref{mgen}) and (\ref{mlen}) we find
\begin{equation}
\begin{array}{cll}
 \mathcal{G}_{0}=g_{00}+\widetilde{\epsilon}_0\mathcal{G}_{0}+V_{0,-2}\mathcal{G}_{-2}
 +V_{0,2}\mathcal{G}_{2}
\end{array}
\end{equation}
\begin{equation}
\begin{array}{cll}
 \quad \quad \quad=g_{00}+\widetilde{\epsilon}_0\mathcal{G}_{0}+V_{0,-2}F_{-2}\mathcal{G}_{0}
 +V_{0,2}B_2\mathcal{G}_{0}
\end{array}
\end{equation}
\begin{equation}
\begin{array}{cll}
\quad \quad =[I-\widetilde{\epsilon}_0-V_{0,-2}F_{-2}-V_{0,2}B_2]^{-1}g_{00}.
\end{array}\label{ghhg0}
\end{equation}
With this we conclude that
\begin{equation}
\begin{array}{cll}
 \mathcal{G}_{-2m}=\left(\prod\limits_{i=1}^{m}F_{-2i}\right)\mathcal{G}_{0},\
 \ m\ge1,
\end{array}
\end{equation}
\begin{equation}
\begin{array}{cll}
 \mathcal{G}_{2m}=\left(\prod\limits_{i=1}^{m}B_{2i}\right)\mathcal{G}_{0},\
 \ m\ge1,
\end{array}
\end{equation}
\begin{equation}
\mathcal{G}_{m,n}(\varepsilon)=\mathcal{G}_{m-n,0}(\varepsilon+n\omega_0/2),
\end{equation}
 where
\begin{equation}
\begin{array}{ll}
 B_{m}=(I-\widetilde{\epsilon}_m-V_{m,m+2}B_{m+2})^{-1}V_{m,m-2},\\
 F_m=(I-\widetilde{\epsilon}_m-V_{m,m-2}F_{m-2})^{-1}V_{m,m+2}.
\end{array}
\end{equation}

\section{The current expressed in terms of Fourier-transformed Green's functions}
Let
$\Gamma_{L}^{\tau}=\gamma\Gamma_{R}^{\tau}$, and $N(t)$ be the
occupation of the central region.
Using the continuity equation
and that the spin channels 
are decoupled
we find for the current
\begin{equation}
\begin{array}{cll}
\hspace*{-2cm}I_L^\tau(t)&=&\frac{1}{1+\gamma}I_L^\tau(t)+\frac{\gamma}{1+\gamma}[e\frac{dN_\tau(t)}{dt}-I_R^\tau(t)]\\

&\hspace*{-3cm}=&\hspace*{-3cm}\frac{\gamma}{1+\gamma}e\frac{dN_\tau(t)}{dt}-\frac{\gamma}{1+\gamma}\frac{e}{\hbar}\Re\sum\limits_{
ss'} \int\frac{d\varepsilon}{2\pi}\int dt_1
e^{-i\varepsilon(t_1-t)}e^{-\frac{is\omega_0
t}{2}}e^{\frac{is'\omega_0 t_1}{2}}\\
&&\hspace*{-3cm}\times\{2\mathcal{G}_{s,s'}^{\tau\tau,r}(t,t_1)
i\Gamma_{R}^{\tau}[f_{L}(\varepsilon)-f_{R}(\varepsilon)]
\}\\

&\hspace*{-3cm}=&\hspace*{-3cm}\frac{\gamma}{1+\gamma}e\frac{dN_\tau(t)}{dt}-\frac{\gamma}{1+\gamma}\frac{e}{\hbar}\Re\sum\limits_{
ss'} \int\frac{d\varepsilon}{2\pi}\int dt_1
e^{-i\varepsilon(t_1-t)}e^{-\frac{is\omega_0
t}{2}}e^{\frac{is'\omega_0 t_1}{2}}\sum\limits_{n}\int\frac{d\varepsilon'}{2\pi}e^{-i\varepsilon'(t-t_1)}e^{in\omega_0(t+t_1)/2}\\
&&\hspace*{-3cm}\times\{2\mathcal{G}_{s,s'}^{\tau\tau,r}(\varepsilon'-n\omega_0/2,\varepsilon'+n\omega_0/2)
i\Gamma_{R}^{\tau}[f_{L}(\varepsilon)-f_{R}(\varepsilon)]
\}\\

&\hspace*{-3cm}=&\hspace*{-3cm}\frac{\gamma}{1+\gamma}e\frac{dN_\tau(t)}{dt}-\frac{\gamma}{1+\gamma}\frac{e}{\hbar}\Re\sum\limits_{n}\sum\limits_{
ss'} \int\frac{d\varepsilon}{2\pi}
\int\frac{d\varepsilon'}{2\pi}\int dt_1 e^{-\frac{is\omega_0
t}{2}}e^{\frac{is'\omega_0 t_1}{2}}e^{-i(\varepsilon'-\varepsilon)(t-t_1)}e^{in\omega_0(t+t_1)/2}\\
&&\hspace*{-3cm}\times\{2\mathcal{G}_{s,s'}^{\tau\tau,r}(\varepsilon'-n\omega_0/2,\varepsilon'+n\omega_0/2)
i\Gamma_{R}^{\tau}[f_{L}(\varepsilon)-f_{R}(\varepsilon)]
\}\\

&\hspace*{-3cm}=&\hspace*{-3cm}\frac{\gamma}{1+\gamma}e\frac{dN_\tau(t)}{dt}-\frac{\gamma}{1+\gamma}\frac{e}{\hbar}\Re\sum\limits_{
ss'}\sum\limits_{n}\int\frac{d\varepsilon'}{2\pi}
\int\frac{d\varepsilon}{2\pi}\int_{-\infty}^{t} dt_1
e^{i(\varepsilon-\varepsilon'+(n-s)\omega_0/2)t}e^{i(-\varepsilon+\varepsilon'+(n+s')\omega_0/2)t_1}\\
\end{array}
\label{current}
\end{equation}
Defining $\varepsilon''=\varepsilon'+\frac{(s'+n)\omega_{0}}{2}$, we may
rewrite the integral over $t_{1}$ as
\begin{equation}
\begin{array}{cl}
&\int_{-\infty}^{t}dt_{1}e^{it_{1}\left(-\varepsilon+\varepsilon'+(n+s')\omega_{0}/2\right)}  \\ =&\int_{-\infty}^{t}dt_{1}\exp[it_{1}(\varepsilon''-\varepsilon)] \\
=&\int_{-\infty}^{\infty}dt_{1}\theta(t-t_{1})\exp[it_{1}(\varepsilon''-\varepsilon)] \\
=&\int_{-\infty}^{\infty}dt_{1}\int\frac{d\omega'}{2\pi i}\frac{\exp[i\omega'(t-t_{1})]}{\omega'-i\eta}\exp[it_{1}(\varepsilon''-\varepsilon)] \\
=&\int\frac{d\omega'}{2\pi i}\frac{\exp(i\omega't)}{\omega'-i\eta}
\underbrace{\int_{-\infty}^{\infty}dt_{1}\exp[it_{1}(\varepsilon''-\varepsilon-\omega')]}_{2\pi\delta(\varepsilon''-\varepsilon-\omega')} \\
=& -i\frac{\exp[i(\varepsilon''-\varepsilon)t]}{\varepsilon''-\varepsilon-i\eta}
=-i\frac{\exp\{i[\varepsilon'-\varepsilon+(s'+n)\omega_{0}/2]t\}}{\varepsilon'-\varepsilon+(s'+n)\omega_{0}/2-i\eta}.
\end{array}
\end{equation}
Upon substituting this expression into Eq. (\ref{current}) we find
\begin{equation}
\begin{array}{cll}
I_{L}^{\tau}(t) &=& \frac{\gamma}{1+\gamma}e\frac{dN_{\tau}(t)}{dt}-\frac{\gamma}{1+\gamma}\frac{e}{\hbar}\Re\sum_{nss'}\exp{(it\omega_{0}[n+(s'-s)/2])} \\
&\times & \int\frac{d\varepsilon}{2\pi}2i\Gamma_{R}^{\tau}[f_{L}(\varepsilon)-f_{R}(\varepsilon)]
\underbrace{\int\frac{d\omega}{2\pi i}\frac{\mathcal{G}_{ss'}^{\tau\tau,r}(\omega-n\omega_{0}/2,\omega+n\omega_{0}/2)}
{\omega-\varepsilon+(n+s')\omega_{0}/2-i\eta}}_{{\rm Integral (II)}}.
\end{array}
\end{equation}
Integral (II) can be evaluated by means of residue method to give
\begin{equation}
\begin{array}{cll}
I_{L}^{\tau}(t) &=& \frac{\gamma}{1+\gamma}e\frac{dN_{\tau}(t)}{dt}-\frac{\gamma}{1+\gamma}\frac{e}{\hbar}\Re\sum\limits_{nss'}
\int\frac{d\varepsilon}{2\pi}2ie^{i(2n-s+s')\omega_{0}t/2}\\
&\times & \left\{\mathcal{G}_{ss'}^{\tau\tau,r}[\varepsilon-(2n+s')\omega_{0}/2,\varepsilon-s'\omega_{0}/2]
\Gamma_{R}^{\tau}[f_{L}(\varepsilon)-f_{R}(\varepsilon)] \right\}.
\end{array}
\end{equation}
%
%
%
The time-averaged current is thus given by
\begin{equation}
\begin{array}{cll}
\langle I_L^\tau\rangle

&=&\frac{\gamma}{1+\gamma}e\left\langle\frac{dN_\tau(t)}{dt}\right\rangle-\frac{\gamma}{1+\gamma}\frac{e}{\hbar}\Re\sum\limits_{
ss'}\sum\limits_{n} \int\frac{d\varepsilon}{2\pi}
\delta_{2n,s-s'}\\
&&\hspace*{-3cm}\times\left\{2\mathcal{G}_{s,s'}^{\tau\tau,r}\left[\varepsilon-(2n+s')\omega_0/2,\varepsilon-s'\omega_0/2\right]
i\Gamma_{R}^{\tau}\left[f_{L}(\varepsilon)-f_{R}(\varepsilon)\right]
\right\}\\

&\hspace*{-3cm}=&\hspace*{-3cm}\frac{\gamma}{1+\gamma}\left\{e\left\langle\frac{dN_\tau(t)}{dt}\right\rangle-\frac{e}{\hbar}\Re\sum\limits_{
ss'}\int\frac{d\varepsilon}{2\pi}
%
2\mathcal{G}_{s,s'}^{\tau\tau,r}(\varepsilon-s\omega_0/2,\varepsilon-s'\omega_0/2)
i\Gamma_{R}^{\tau}[f_{L}(\varepsilon)-f_{R}(\varepsilon)]\right\}\\

&=&-\frac{\gamma}{1+\gamma}\frac{ie}{\hbar}\sum\limits_{
ss'}\int\frac{d\varepsilon}{2\pi}\Gamma_{R}^{\tau}[f_{L}(\varepsilon)-f_{R}(\varepsilon)]
\\
&&\times
\left[\mathcal{G}_{s,s'}^{\tau\tau,r}\left(\varepsilon-s\omega_0/2,\varepsilon-s'\omega_0/2\right)
-\mathcal{G}_{s,s'}^{\tau\tau,a}\left(\varepsilon-s\omega_0/2,\varepsilon-s'\omega_0/2\right)\right]
 \\

&=&-\frac{\gamma}{1+\gamma}\frac{ie}{\hbar}\sum\limits_{
ss'}\int\frac{d\varepsilon}{2\pi}
  \left[\mathcal{G}_{-s,-s';ss'}^{\tau\tau,r}(\varepsilon)
-\mathcal{G}_{-s,-s';ss'}^{\tau\tau,a}(\varepsilon)\right]
\Gamma_{R}^{\tau}\left[f_{L}(\varepsilon)-f_{R}(\varepsilon)\right]
 \\

 &=&-\frac{ie}{\hbar}\sum\limits_{
ss'}\int\frac{d\varepsilon}{2\pi}
  \left[\mathcal{G}_{-s,-s';ss'}^{\tau\tau,r}(\varepsilon)
-\mathcal{G}_{-s,-s';ss'}^{\tau\tau,a}(\varepsilon)\right]
\frac{\Gamma_{L}^{\tau}\Gamma_{R}^{\tau}}{\Gamma_{L}^{\tau}+\Gamma_{R}^{\tau}}[f_{L}(\varepsilon)-f_{R}(\varepsilon)]
. \\
\end{array}
\end{equation}
With this we conclude that
\begin{equation}
\begin{array}{cll}
\langle I_L\rangle &=&\langle I_L^\uparrow\rangle+\langle I_L^\downarrow\rangle\\

&=&-\frac{ie}{\hbar}\sum\limits_{
ss'\tau}\int\frac{d\varepsilon}{2\pi}
  \left[\mathcal{G}_{-s,-s';ss'}^{\tau\tau,r}(\varepsilon)
-\mathcal{G}_{-s,-s';ss'}^{\tau\tau,a}(\varepsilon)\right]
\frac{\Gamma_{L}^{\tau}\Gamma_{R}^{\tau}}{\Gamma_{L}^{\tau}+\Gamma_{R}^{\tau}}[f_{L}(\varepsilon)-f_{R}(\varepsilon)]
 ,\\
\end{array}
\end{equation}

For numerical implementation we use the expression (we employ $\Gamma_L=\Gamma_R$, i.e. $\gamma=1$)
\begin{equation}
\begin{array}{cll}
\langle I_L\rangle &=&-\frac{ie}{2h}\sum\limits_{\tau}\int
d\varepsilon \left\{
  [\mathcal{G}_{-1,-1;++}^{\tau\tau,r}(\varepsilon)
-\mathcal{G}_{-1,-1;++}^{\tau\tau,a}(\varepsilon)]

+[\mathcal{G}_{-1,1;+-}^{\tau\tau,r}(\varepsilon)
-\mathcal{G}_{-1,1;+-}^{\tau\tau,a}(\varepsilon)]  \right.\\

&&\hspace*{-3cm}+\left.[\mathcal{G}_{1,-1;-+}^{\tau\tau,r}(\varepsilon)
-\mathcal{G}_{1,-1;-+}^{\tau\tau,a}(\varepsilon)]

+[\mathcal{G}_{1,1;--}^{\tau\tau,r}(\varepsilon)
-\mathcal{G}_{1,1;--}^{\tau\tau,a}(\varepsilon)]\right\}
\Gamma_{R}^{\tau}\left[f_{L}(\varepsilon)-f_{R}(\varepsilon)\right]
 \\

&\hspace*{-3cm}=&\hspace*{-3cm}-\frac{ie}{2h}\sum\limits_{\tau}\int d\varepsilon\{
  [\mathcal{G}_{00;++}^{\tau\tau,r}(\varepsilon-\frac{\omega_0}{2})
-\mathcal{G}_{00;++}^{\tau\tau,a}(\varepsilon-\frac{\omega_0}{2})]

+[\mathcal{G}_{-20;+-}^{\tau\tau,r}(\varepsilon+\frac{\omega_0}{2})
-\mathcal{G}_{-20;+-}^{\tau\tau,a}(\varepsilon+\frac{\omega_0}{2})]\\

&&\hspace*{-3cm}+[\mathcal{G}_{20;-+}^{\tau\tau,r}(\varepsilon-\frac{\omega_0}{2})
-\mathcal{G}_{20;-+}^{\tau\tau,a}(\varepsilon-\frac{\omega_0}{2})]

+[\mathcal{G}_{00;--}^{\tau\tau,r}(\varepsilon+\frac{\omega_0}{2})
-\mathcal{G}_{00;--}^{\tau\tau,a}(\varepsilon+\frac{\omega_0}{2})]\}\\
&&\hspace*{-3cm}\times \Gamma_{R}^{\tau}[f_{L}(\varepsilon)-f_{R}(\varepsilon)].

\end{array}
\end{equation}

\end{document}